\documentclass[10pt,twocolumn,twoside]{IEEEtran} 
\usepackage{graphics} 
\usepackage{epsfig} 
\usepackage{mathptmx} 
\usepackage{times} 
\usepackage{amsmath} 
\usepackage{amssymb}  
\usepackage{xcolor}
\usepackage[lofdepth,lotdepth]{subfig}
\DeclareMathOperator*{\esssup}{ess\,sup}
\usepackage{epstopdf}
\usepackage[para]{threeparttable}
\newtheorem{theorem}{Theorem}[section]
\newtheorem{lemma}[theorem]{Lemma}
\newtheorem{proposition}[theorem]{Proposition}

\newtheorem{definition}[theorem]{Definition}
\newtheorem{assumption}[theorem]{Assumption}
\newtheorem{remark}[theorem]{Remark}

\newtheorem{example}[theorem]{Example}
\ifCLASSINFOpdf
\else
\fi
\begin{document}

\title{\LARGE \bf Self-triggering in Vehicular Networked Systems with State-dependent Bursty Fading Channels}
%
%
%

\author{Bin Hu and Michael Lemmon
\thanks{This work was supported by the National Science
	Foundation under Grant CNS-1239222.}
\thanks{Bin Hu is with Department of Engineering Technology, Old Dominion University, Norfolk, VA, 23529, US. {\tt\small bhu@odu.edu}}
\thanks{Michael Lemmon is with Department of Electrical Engineering, University of Notre Dame, Notre Dame, IN, 46556, US. {\tt\small lemmon@nd.edu}} }

\maketitle

\begin{abstract}
Vehicular Networked Systems (VNS) are mobile ad hoc networks where vehicles exchange information over wireless communication networks to ensure safe and efficient operation. It is, however, challenging to ensure system safety and efficiency as the wireless channels in VNS are often subject to \emph{state-dependent deep fades} where the data rate suffers a severe drop and changes as a function of vehicle states. Such couplings between vehicle states and channel states in VNS thereby invalidate the use of separation principle to design event-based control strategies. By adopting a \emph{state-dependent fading channel model} that was proposed to capture the interaction between vehicle and channel states, this paper presents a novel self-triggered scheme under which the VNS ensures efficient use of communication bandwidth while preserving stochastic stability. The novelty of the proposed scheme lies in its use of the state-dependent fading channel model in the event design that enables an adaptive and effective adjustment on transmission frequency in response to dynamic variations on channel and vehicle states. Under the proposed self-triggered scheme, this paper presents a novel source coding scheme that tracks vehicle's states with performance guarantee in the presence of state-dependent fading channels. The efficacy and advantages of the proposed scheme over other event-based strategies are verified through both simulation and experimental results of a leader-follower example.
\end{abstract}


%
\IEEEpeerreviewmaketitle

\section{Introduction}
\subsection{Background and Motivation}
Vehicular Networked Systems (VNS) consist of numerous vehicles coordinating their operations by exchanging information over a wireless radio communication network. VNS represents one type of mobile ad hoc networks that have been deployed in a variety of safety-critical applications, such as intelligent transportation systems with Vehicle to Vehicle~(V2V) communication \cite{papadimitratos2009vehicular, kenney2011dedicated, cheng2007mobile}, air transportation systems with Automatic Dependent Surveillance-Broadcast (ADS-B) \cite{park2014high, park2012investigating}  and underwater autonomous vehicles with optic or acoustic communication \cite{akyildiz2005underwater, partan2007survey}.  

These vehicular networks, however, are often subject to \emph{deep fades} where the data rate drops precipitously and remains low over a contiguous period of time. Such \emph{deep fades} functionally depend on the vehicles' physical states (e.g. inter-vehicle distance, velocities and heading angles)~\cite{tse2005fundamentals,cheng2007mobile}. Deep fades inevitably cause a significant degradation in the overall vehicular system performance and result in undesirable safety issues, such as vehicle collisions. To maintain system safety and quality performance, VNS may require a large amount of communication resources, such as channel bandwidth, to recover the performance loss due to deep fades. The overuse of communication resources, however, will certainly compromise the long-term operational normalcy of the system since electronic devices have limited energy. Therefore, both the safety and efficient use of communication resources in the system must be jointly  evaluated to ensure a successful implementation of VNS.

Recent studies have shown that an event-based communication scheme is an effective approach to reduce the required communication bandwidth to maintain a specified system performance \cite{wang2011event}. In an event-triggered communication framework, the transmission of information only occurs when the novelty of system states exceeds a predefined state-dependent threshold. Recent work has shown that the system performance under such a state-dependent triggering scheme can be preserved when the communication delay or the number of consecutive packet loss is bounded \cite{guinaldo2012distributed, wang2011event}. Such robustness, however, can be easily violated in VNS where a burst of delay or packet dropouts may occur with a nonzero probability. To address this issue, our prior work~\cite{hu2014event} proposed a new event-triggered scheme that ensures \emph{almost sure stability} for VNS with a bursty fading channel. By exploring the dependence of channel conditions on vehicular states, it is shown that the transmission time interval generated by the event-triggered scheme increases monotonically as the system state approaches its equilibrium, thereby providing an efficient use of channel bandwidth. 

This paper extends our previous results in \cite{hu2014event} in three nontrivial aspects. First, this paper proposes a stochastic hybrid system framework that generalizes the VNS structure considered in the prior work \cite{hu2014event}. Such generalization on the system framework enables the results of this paper to apply to a variety of realistic vehicular applications, such as the vehicle platoon in \cite{ploeg2014lp}, the leader-follower mobile robotic system in \cite{tanner2004leader} and the air traffic control system in \cite{park2014high, tomlin2000game}.  Secondly, by relaxing the uniformly Lipschitz assumption on the system dynamics in our former results \cite{hu2014event}, this paper develops a more general self-triggered  and encoder/decoder scheme under which four types of \emph{stochastic stability} are ensured for VNS. Thirdly, extensive simulation as well as experimental results are presented in this paper to further demonstrate the benefits and advantages of our proposed self-triggered scheme under bursty fading channels against traditional event-triggered schemes, such as \cite{wang2011event, wang2009self, tabuada2007event}.
\subsection{Related Work and Contributions}
The research topics on event-based communication and control have attracted a great deal of attention in both the control and communication research communities \cite{hetel2017recent}. It is beyond the scope of this paper to do an exhaustive literature review on this popular topic. Instead, this section will focus on discussing the relationship and connection between our proposed work and the existing research work on event/self triggered schemes under unreliable communication. For those who are interested in more complete review on self-triggered control and event-triggered control, please see \cite{heemels2012introduction} for a recent overview.

The stability and performance of event-based strategies must be carefully examined in the presence of unreliable wireless communication. Prior work in \cite{walsh2002stability, zhang2001stability,lehmann2012event, wu2014formal} have shown that the temporal communication failures caused by packet loss or delay may lead to stability issues for networked control systems. To address such issues in networked control systems with event-based strategies, a great deal of research work \cite{dolk2017event, yu2013event, wang2011event, wang2009self, guinaldo2012distributed, peng2013event, dolk2017output}  have been devoted to find sufficient conditions on communication performance measures, such as the maximum allowable number of successive packet drops~(MANSD) or maximum allowable delay, under which the system stability and performance criteria, such as $\mathcal{H}_{\infty}$ \cite{peng2013event} and $\mathcal{L}_{p}$ \cite{yu2013event, wang2009self, dolk2017output}, can be preserved under the event/self-triggered strategies. 

Two key assumptions for these prior results include that (1) wireless communication channels must be sufficiently reliable to strictly satisfy the MANSD or maximum allowable delay, and (2) the variations on wireless channels must be decoupled from dynamics of the vehicle systems. These assumptions, however, will mostly likely fail to hold for VNS because the wireless channel in VNS is subject to deep fades and may produce a burst of packet loss with a non-zero probability. In addition, as shown in both theoretical and experimental results \cite{cheng2007mobile, akki1994statistical,akki1986statistical}, the channel conditions in VNS are highly dependent on the physical states of the vehicle, such as the inter-vehicle distance, velocities and heading angles.  Such correlation between vehicle and channel states clearly invalidate the use of separation principle in event-triggered design, which is widely adopted in the existing literature \cite{borgers2014event,li2016event}. 

Another challenge of using the event-based strategies under unreliable communication is that a strictly positive \emph{minimum inter-event time} (MIET) may not be guaranteed if packet loss or delay is present. As discussed in \cite{mazo2008event}, the loss of MIET inevitably leads to Zeno phenomenon that generates infinite transmissions or samplings with a finite time interval and seriously hinders the practical implementation of event-based strategies. The main approach to address the potential Zeno issues is a combined framework of time-triggered and event-triggered strategies where the event-based scheme is designed based on a predefined equidistant time instances~(e.g., periodic event-triggered scheme proposed in \cite{peng2013event, heemels2013periodic, antunes2014rollout}) and switched to a time-triggered scheme if packet loss or delay occurs \cite{lehmann2012event, guinaldo2012distributed}. Although the MIET can be always guaranteed to be positive under the combined framework of event-triggered and time-triggered schemes, it is unclear, however, how efficient and effective such combined approaches may be in deep fading channels where a long string of consecutive packet loss may occur.   

Motivated by the challenges discussed above,  the objective of this paper is to design a new self-triggered communication scheme that ensures both stability of VNS and efficient use of communication resources by taking into account the state-dependency and burstiness properties. The key difference between the proposed self-triggered scheme in this paper and the others in the literature, such as \cite{anderson2015self, wang2009self,gommans2014self,anderson2015self} lies in two aspects. First,  by adopting a state-dependent bursty fading channel model proposed in \cite{bin2013, 2015Hu}, this paper explicitly incorporates the knowledge of correlations between communication channel and vehicle states into the design process, which allows the self-triggered scheme to adaptively adjust the transmission frequency in response to any changes in channel conditions. This paper also shows that the inclusion of such correlation knowledge provided by the state-dependent channel model is essential for the proposed self-triggered scheme to achieve efficient utilization of communication bandwidth. Second, unlike the combined framework that relies on a pre-selected minimum time interval to ensure a positive MIET,  the proposed self-triggered scheme guarantees Zeno-free transmission behavior in the presence of bursty-fading channels while still preserving specified system performance. In addition, this paper demonstrates the communication efficiency of the proposed self-triggered scheme through extensive simulation results that compare the communication performance under our proposed self-triggered scheme, such as minimum transmission time interval and average transmission time interval, against that under other existing event-based strategies \cite{wang2011event, li2017efficiently}.  

To summarize, the main contribution of the present paper is a novel self-triggered strategy for VNS in the presence of state-dependent bursty fading channels, ensuring both stochastic stability and efficient use of communication resources. The technical contributions of this paper are summarized below:
\begin{itemize}
	\item This paper proposes a stochastic hybrid system framework under which the self-triggered strategy proposed in this paper can be applied to a variety of VNS including the vehicle platoon systems\cite{ploeg2014lp}, leader-follower mobile robotic systems\cite{tanner2004leader} and air traffic control systems\cite{park2014high}.
	\item With the powerful stochastic hybrid system framework, this paper develops self-triggered strategies under which four different types of stochastic stability are ensured for VNS while respecting the Zeno-free conditions. To the best of our knowledge, this is the first set of results that demonstrate how to design event-based strategies for VNS with state-dependent bursty fading channels.
	\item Under the proposed self-triggered scheme, this paper presents a novel  source coding scheme that is able to track vehicle's state information in the presence of a time varying data rate. In addition, this paper shows that the encoder/decoder can be explicitly constructed such that the tracking performance can be guaranteed even under a time varying data rate that is stochastically changed as a function of the vehicle state.
	\item The advantages of the proposed self-triggered scheme in efficient use of communication bandwidth and preserving system performance over other event-based strategies are demonstrated through both simulation results in Matlab and experimental results in MobileSim robot simulator with real parameters of Pioneer 3-DX mobile robots \cite{Mobilesim}. 
\end{itemize} 

The rest of this paper is organized as follows. Section \ref{section: problem-formulation} describes the models of vehicle dynamics, wireless communication and control systems. Based on the system models presented in Section \ref{sec: main-results}, Section \ref{section: problem-formulation} provides formal definitions of stochastic hybrid system framework as well as stability stabilities. Section \ref{sec:assumption} discusses the necessary assumptions needed to establish the main results of this paper. With the assumptions stated in Section \ref{sec:assumption}, Section \ref{sec: main-results} states the main results of this paper. The main results are applied to a leader-follower tracking example presented in Section \ref{sec: leader-follower} and verified through both simulation and experimental results provided in Section \ref{sec: simulation}. Section \ref{section: conclusion} concludes the paper. 

{\textbf{Notations}}.
Let $\mathbb{R}^{n}$ denote the $n$-dimensional Euclidean vector space, and $\mathbb{R}_{+}$, $\mathbb{Z}_{+}$ denote the nonnegative reals and integers respectively. The infinity norm of the vector $x \in \mathbb{R}^{n}$ is denoted by $|x|:=\max_{1 \leq i \leq n}{|x_{i}|}$ where $x_{i}$ is the $ith$ element of the vector $x$. Consider the real valued function $x(\cdot): \mathbb{R}_{+} \rightarrow \mathbb{R}^{n}$,  $x(t)$ denotes the value that function $x(\cdot)$ takes at time $t \in \mathbb{R}_{+}$. The left limit value of $x$ at time $t$ is denoted by $x(t^{-})$. Given a time interval $[t_{1}, t_{2})$ with $t_{1}, t_{2} \in \mathbb{R}_{+}$, the essential supremum of the function $x(t)$ over a time interval $[t_{1}, t_{2})$ is denoted as $|x(t)|_{[t_{1}, t_{2})}=\esssup_{t \in [t_{1}, t_{2})}\|x(t)\|$ where $\|x(t)\|$ is the Euclidean norm of function $x$ at time $t$. The function $x(t)$ is essentially bounded if there exists a positive real $M < \infty$ such that $|x(t)|_{\mathcal{L}_{\infty}}=\esssup_{t > 0}\|x(t)\| \leq M$.

A function $\alpha(\cdot): \mathbb{R}_{+} \rightarrow \mathbb{R}_{+}$ is a class $\mathcal{K}$ function if it is continuous and strictly increasing, and $\alpha(0)=0$. $\alpha(t)$ is a class $\mathcal{K}_{\infty}$ function if it is class $\mathcal{K}$ and radially unbounded. A function $\beta(\cdot, \cdot): \mathbb{R}_{+} \times \mathbb{R}_{+} \rightarrow \mathbb{R}_{+}$ is a class $\mathcal{KL}$ function if $\beta(\cdot, t)$ is class $\mathcal{K}_{\infty}$ function for each fixed $t \in \mathbb{R}_{+}$ and $\beta(s, t) \rightarrow 0$ for each $s \in \mathbb{R}_{+}$ as $t \rightarrow +\infty$. $\beta(\cdot, \cdot)$ is an exp-$\mathcal{KL}$ function if there exists positive reals $c_{i} \in \mathbb{R}_{+}, i=1,2$ such that $\beta(s, t)=c_{1}\exp(-c_{2}t)s$.
\section{System Description}
\label{section: system-description}
\begin{figure}[t]
\centering
	\includegraphics[width=0.45\textwidth]{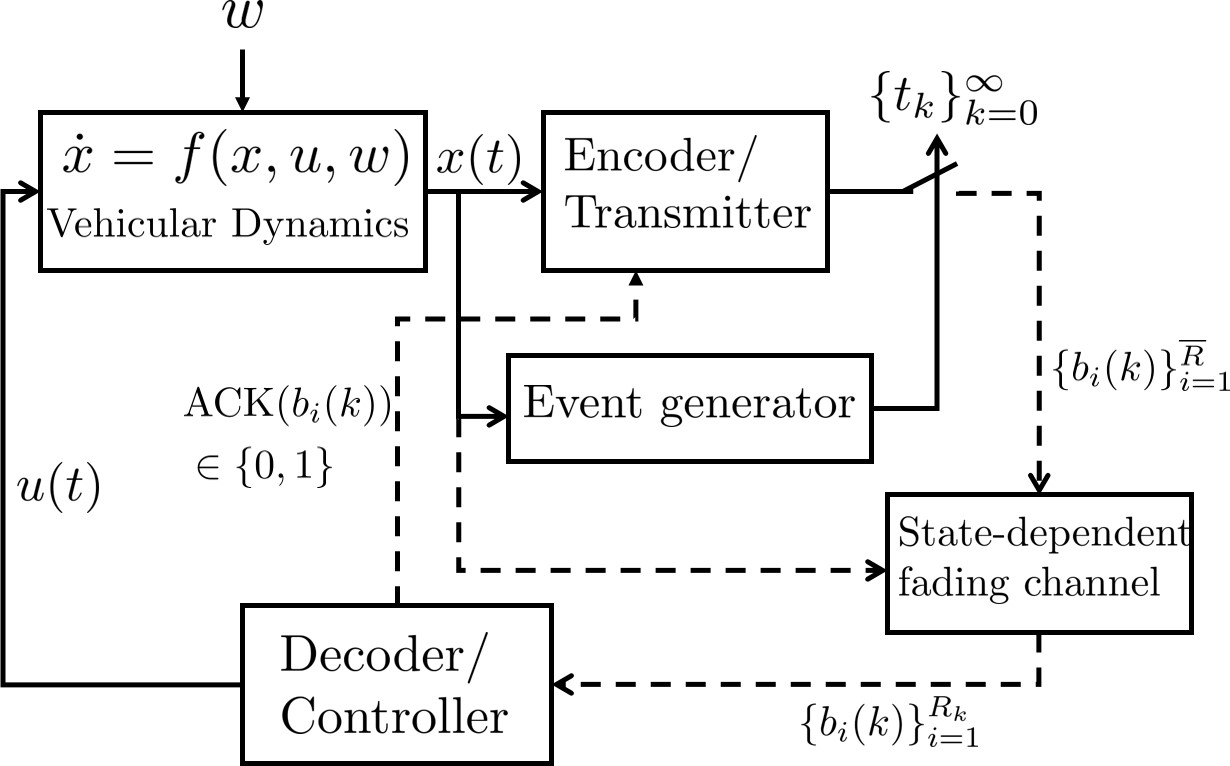}
	\caption{Self-triggered Vehicular Networked System}
	\label{fig:structure}
\end{figure}

Fig.~\ref{fig:structure} shows a self-triggered control framework for a vehicular networked system that consists of blocks of \emph{Vehicular Dynamics}, \emph{Encoder/Transmitter}, \emph{Event generator}, \emph{State-dependent fading channel} and \emph{Decoder/Controller}. The following subsections focus on the detailed descriptions of these blocks.

\subsection{Vehicular Dynamics}
Consider that the dynamics of a vehicular system satisfy the following nonlinear ODE, 
\begin{align}
\dot{x}=f(x, u, w), \quad x(0)=x_{0}
\label{eq:sys}
\end{align}
where $x \in \mathbb{R}^{n}$ is the system state that may represent the inter-vehicle distance and relative bearing angles (see Section \ref{sec: leader-follower} in this paper or \cite{hu2014event}), $u \in \mathbb{R}^{m}$ is the control input to the system and $w \in \mathbb{R}^{\ell}$ denotes the external disturbance that is essentially ultimately bounded, i.e. $\exists W > 0, \text{s.t.} \, |w|_{\mathcal{L}_{\infty}} \leq W$. The vector field~$f(\cdot, \cdot, \cdot): \mathbb{R}^{n} \times \mathbb{R}^{m} \times \mathbb{R}^{\ell} \rightarrow \mathbb{R}^{n}$ is a locally Lipschitz function.

The control objective for vehicular networked systems is to track predefined set-points in the presence of bursty fading channels. 
The tracking performance is investigated under two communication constraints: (1) State measurements $x(t)$ are taken and only available to controller at discrete time instants $t_k \in \mathbb{R}_{\geq 0}, k \in \mathbb{Z}_{\geq 0}$; (2) The sampled state measurements $x(t_k)$ used for tracking control, are encoded by a finite number of symbols and subject to stochastically varying data rates.

\subsection{Event-based Communication: Encoder/Transmitter and Event Generator}
The continuous vehicular state $x(t)$ in Fig.~\ref{fig:structure} is sampled at discrete time instants $\{t_k\}_{k=0}^{\infty}$ with $t_{k} < t_{k+1}$ and $t_{k} \in \mathbb{R}_{+}, \forall k \in \mathbb{Z}_{+}$. Such strictly increasing time instants $\{t_k\}_{k=0}^{\infty}$ are generated by an \emph{Event generator}, which decides when to transmit state information. The sampled state $x(t_k)$ at time instant $t_k$ is quantized by an \emph{Encoder} with $\overline{R}$ blocks of bits $\{b_{i}(k)\}_{i=1}^{\overline{R}}$. Each block consists of $n$ binary bits, which characterize the information of the states. Thus, the continuous vehicular state at each \emph{discrete} time instant $t_k$ will be encoded and represented by one of the $2^{n\overline{R}}$ finite symbols. We assume that the symbol with $\overline{R}$ blocks of bits are assembled into $\overline{R}$ number of small packets with a packet length $n$, and sequentially transmitted across a wireless fading channel. In this paper, we assume that the time spent on quantization and packet-assembly is sufficiently small and its impact on system stability and performance can be safely neglected. 

\begin{figure}[t]
	\centering
	\includegraphics[width=0.45\textwidth]{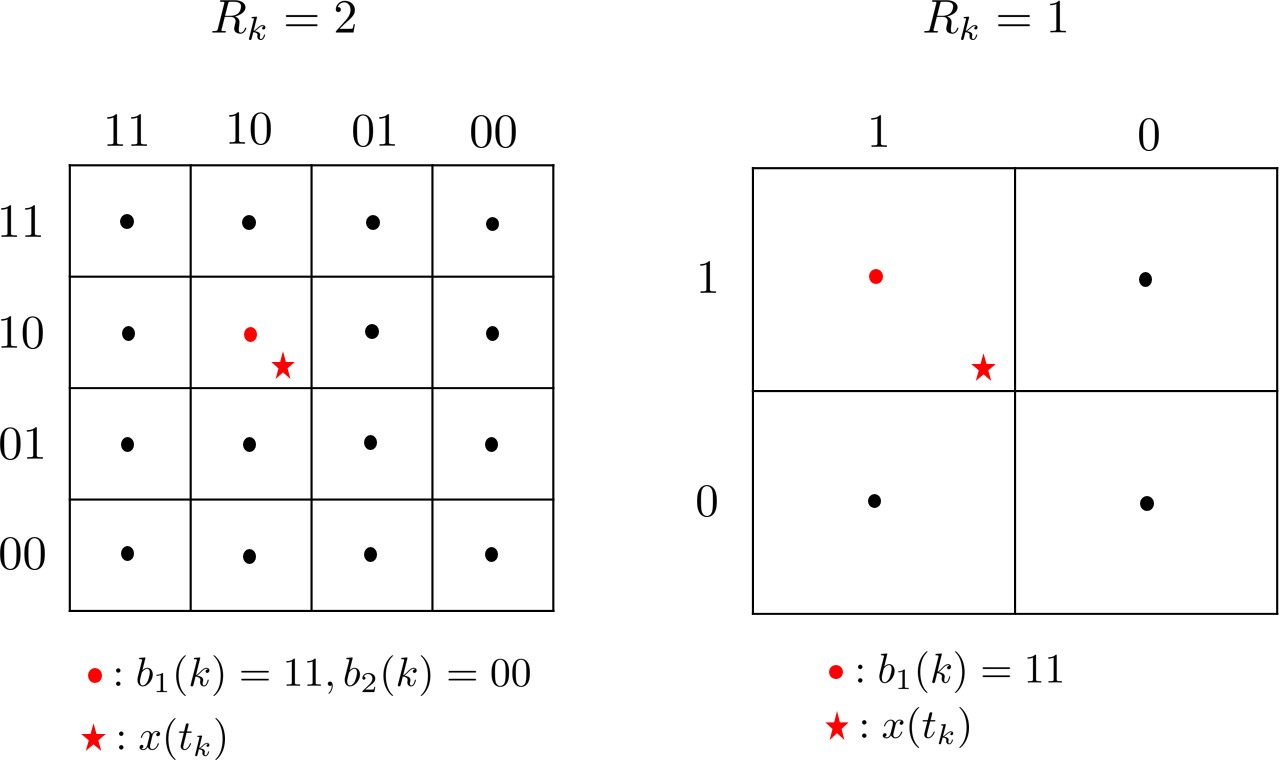}
	\caption{Box based Quantizer with different number of received bits.}
	\label{fig:quantizer}
\end{figure}

\textbf{Sequential Transmission in VNS.} Unlike most stationary or slow varying wireless network, the wireless channels in vehicular systems often exhibit much faster variations due to high motions in vehicular transceivers \cite{cheng2007mobile}. Recent work has shown that vehicular wireless channels, such as V2V communication, are subject to small coherence time, which makes the transmission of a large size of packets fairly challenging. Motivated by this challenge, a sequential communication scheme is adopted in this paper to sequentially transmit prioritized small packets over wireless channels \cite{kenney2011dedicated, hu2015distributed, papadimitratos2009vehicular}. Specifically, the sequential transmission scheme ensures that packets with the highest priority~(most significant bits) are received first \cite{martins2006feedback}. In comparison to the transmission policy that wraps all information into one single big packet, the sequential transmission protocol with small prioritized packets is able to recover the transmitted information with a reduced accuracy in the presence of bursty fading channels. Example \ref{example:small-packets} illustrates the basic ideas of sequential transmission with prioritized small packets.
\begin{example}
	\label{example:small-packets}	
	Consider a two-dimensional vehicle state $x \in \mathbb{R}^{2}$, and suppose $\overline{R}=2$ for each dimension. Let $R_{k} \leq \overline{R}$ denote the number of blocks of bits successfully received at the \emph{Decoder/Controller} side at time instant $t_k, \forall k \in \mathbb{Z}_{\geq 0}$.  Fig. \ref{fig:quantizer} shows the cases of receiving either one~(plot on the right-hand side) or two~(plot on the left-hand side) blocks of bits under the box based quantizer. Specifically, the quantizer first partitions the region~(big square) that contains the true value of vehicle state $x(t_k)$ marked by the red star, into $2^{2R_{k}}$ equidistant small squares. The vehicle state $x(t_k)$ is then estimated by the center~(marked by the red dot) of the small square that contains the true value of vehicle state $x(t_k)$. Such an estimate is encoded by binary sequences $\{b_{i}(k)\}_{i=1}^{R_{k}}$ that are assembled into $R_{k}$ number of packets. The packet with the most significant bits, e.g. $b_{1}(k)=11$ in Fig. \ref{fig:quantizer}, is given the highest priority. The packet priority decreases as the integer $i$ increases. Thus, with $\overline{R}=2$, two packets with $b_{1}(k)=11$ and $b_{2}(k)=00$ are generated by the \emph{Encoder} and are transmitted over fading channels. Suppose the \emph{Decoder/Controller}  and the \emph{Encoder} has the same codebook as shown in Fig. \ref{fig:quantizer}, the decoder can reconstruct the same state estimate if two packets are successfully received. If only one packet is received, i.e., the most significant bits $b_{1}(k)=11$ in this example, the state's estimate can still be reconstructed but with a reduced precision as shown in the right-hand-side plot in Fig. \ref{fig:quantizer}. 
\end{example}

\subsection{State-dependent V2V Fading Channel}
The number of successfully received packets $R_{k}$ at each transmission time instant $t_{k}$ randomly changes due to bursty fading in vehicular channels.  This paper adopts a state-dependent exponential bounded burstiness (SD-EBB) model to characterize the stochastic variations on $R_{k}$ \cite{bin2013}. The SD-EBB models were developed in our recent work \cite{2015Hu}, and were able to describe a wide range of fading channels including i.i.d. and Markov chain channels. More importantly, the SD-EBB model explicitly characterize the probability bound on channel burstiness and its dependency on vehicle state, which has been proven to be essential for system stability \cite{2015Hu, bin2013}. To be specific, let $h(\cdot)$ and $\gamma(\cdot)$ denote continuous, nonnegative, monotone decreasing functions from $\mathbb{R}_{+}$ to $\mathbb{R}_{+}$. Assume that the probability of successfully receiving $R_{k}$ packets at time instant $t_k$ satisfies
\begin{align}
\label{ineq:EBB}
{\rm Pr}\{R_{k} \leq h(|x(t_{k})|)-\sigma\} \leq e^{-\gamma(|x(k)|)\sigma}
\end{align}
with $\sigma \in [0, h(|x(t_k)|)]$. The function $h(|x(t_k)|)$ in SD-EBB model is a state-dependent threshold that separates the low bit-rate region from the high bit rate region in the channel state space. It monotonically decreases as vehicle states~(e.g. inter-vehicle separation and relative bearing angle) deviates from the origin. The state-dependent function $h(|x(t_k)|)$ models the impact of large scale fading caused by path loss and directional antenna gain on data rate \cite{choudhury2002using,tse2005fundamentals}. The variable $\sigma \in [0, h(|x(t_k)|)]$ is the \emph{dropout burst length} in the low bit-rate region. Thus, the left hand side of the SD-EBB model characterizes the probability of fading channels exhibiting a bursty packet loss with a burst length $\sigma$. The right-hand side of the SB-EBB model shows that such a bursty probability is exponentially bounded. The function $\gamma(|x(t_k)|)$ is a state-dependent exponent in the probability bound that characterizes how fast the probability of
a bursty dropout decays as a function of dropout burst length within the low bit rate region. 

The SD-EBB model can be used in a variety of vehicular applications, such as leader-follower formation control for ground transportation system \cite{cheng2007mobile}, air transportation systems \cite{park2014high} and autonomous underwater vehicles \cite{akyildiz2005underwater}, where large inter-vehicle distance and vehicular velocities cause low data rate and more likely lead to deep fades, or in ad hoc wireless networks with directional antennae where changes in the relative bearing angle between the transmitter and receiver may cause a deep fade \cite{yi2003capacity}. Example \ref{example:SD-EBB} shows that how the SD-EBB model is obtained and related to the notion of \emph{outage probability} which is well-known performance metric for fading channels. 

\begin{example}
	\label{example:SD-EBB}
Let $X_{i}(k) \in \{0, 1\}$ denote a binary random variable at time instant $t_k$, with $X_{i}(k)=1$ representing the successful reception of the $i^{th}$ block of bits~(packet) and $X_{i}(k)=0$ otherwise, then $R_{k}=\sum_{i=1}^{\overline{R}}X_{i}(k)$. For a given transmission power $p$ and threshold $\gamma_0$, one has
\begin{align}
{\rm Pr}\{X_{i}(k)=1\}&=1-{\rm Pr}\{\text{SNR} \leq \gamma_0 \} \nonumber\\
&=1-{\rm Pr}\{pg^{2}/(\vartheta(x(t_k))N_{0}) \leq \gamma_0\} \nonumber\\
&\overset{\bigtriangleup}{=} \varpi(x(t_k)),
\label{eq:probability-reception}
\end{align}
where $N_0$ is the noise power, $g$ is a random variable that characterizes the small scale fading, and $\vartheta(x(t_k))$ is a continuous, positive and monotonically increasing function that characterizes the path loss and directional antenna gain in wireless channels, e.g., $\vartheta(x)=\cos{\alpha}/L^{\nu}$ with path loss exponent $\nu \in [2, 4]$ and vehicle state $x=[L;\alpha]$ where $L$ is the distance between transmitter and receiver, and $\alpha$ is the bearing angle of the directional antenna \cite{balanis2016antenna, stuber2011principles}. $\varpi(|x(t_k)|)$ is the successful reception probability for the $i^{th}$ block of bits~(packet) whose value increases as the vehicle state $x$ moves toward the origin. With the probability in \eqref{eq:probability-reception}, one can obtain the SD-EBB characterization in \eqref{ineq:EBB} by using the Chernoff inequality \cite{hu2015distributed}.	
\end{example}
\subsection{Remote Tracking Control System under Event-based Dynamic Quantization}
The control objective for vehicular networked systems is to track predefined set-points. Rather than using the true value of the vehicle states, the controller has only access to the estimate of the vehicle state that is constructed by an \emph{encoder/decoder} pair. As discussed above, the estimation error induced by the wireless network is highly dependent on the number of packets successfully received at each transmission time instant. By using the SD-EBB model, this section discusses a model-based tracking control system under a dynamic quantization method. 

Let $x^{d} \in \mathbb{R}^{n}$ denote a desired constant set-point that is known to both encoder and decoder ahead of time. Then,  $\hat{\overline{x}} \triangleq \hat{x}-x^{d}$ is the estimate of the vehicle state. In between the transmission time instants $t_{k}, k \in \mathbb{Z}_{\geq 0}$, the state estimate $\hat{\overline{x}}$ and the control action $u(t)$ are generated as follows,
\begin{align}
\label{eq:controller}
\dot{\hat{\overline{x}}}&=f(\hat{\overline{x}}+x^{d}, \kappa(\hat{\overline{x}}), 0) \nonumber \\
u &=\kappa(\hat{\overline{x}}), \quad \forall t \in [t_{k}, t_{k+1})
\end{align}
where $\kappa(\cdot): \mathbb{R}^{n} \rightarrow \mathbb{R}^{m}$ is a nominal feedback control law  ensuring that the estimate state $\hat{\overline{x}}$ in the tracking control system \eqref{eq:controller} asymptotically converges to zero. At each transmission time instant $t_k$, the state estimate $\hat{\overline{x}}(t_k^{+})$ is reset to be a new value obtained from the quantized measurement of a \emph{dynamic quantizer}. The dynamic quantizer in the \emph{Encoder/Decoder} pair is defined by three parameters, $R_{k} \in \mathbb{Z}_{\geq 0}$~(a random variable that defines the number of blocks of bits received at time instant $t_k$), $\hat{\overline{x}}(t_{k})$~(state estimate at time instant $t_k$) and $U(t_k)$~(an auxiliary variable that defines the size of the quantization regions at time instant $t_k$). Consider a box dynamic quantizer and let $\hat{\overline{x}}(t_k)$ denote the center of a hypercubic box with edge length $2U(t_k)$, as shown in Fig. \ref{fig:quantizer}, the quantizer divides the hypercubic box into $2^{nR_{k}}$ equal smaller sub-boxes after receiving $R_{k}$ number of blocks of bits. The sub-box that contains the true value of vehicle state $\overline{x}(t_k)$ is encoded by $\{b_{i}(k)\}_{i=0}^{R_{k}}$. Thus, the new state estimate $\hat{\overline{x}}(t_{k}^{+})$ after receiving the symbol $\{b_{i}(k)\}_{i=0}^{R_{k}}$ can be updated according to the following jump equation of the form 
\begin{align}
\label{eq: quantizer-1}
\hat{\overline{x}}(t_{k}^{+})&=g_{\overline{x}}\bigg(T_{k}, \{b_{i}(k)\}_{i=0}^{R_{k}}, \hat{\overline{x}}(t_k), U(t_k)\bigg), \\
U(t_{k}^{+})&=U(t_k)2^{-R_{k}} 
\label{eq: quantizer-2}
\end{align}
with $\hat{\overline{x}}(t_{k}^{+})$ being the center of a new hypercubic box with an updated edge length $2U(t_{k}^{+})$. Let $T_{k} \triangleq t_{k}-t_{k-1}$ denote a time interval for the $k^{th}$ transmission. Until the $(k+1)^{th}$ transmission, i.e., time instant $t_{k+1}$, the size of the quantization region $U(t_{k+1})$ is propagated according to
\begin{align}
\label{eq: quantizer-3}
U(t_{k+1})=g_{U}(T_{k+1}, \hat{\overline{x}}(t_k^{+}), U(t_{k}^{+})).
\end{align}
and then the procedure of \eqref{eq: quantizer-1}-\eqref{eq: quantizer-3} is repeated.

To enable the feasibility of the event-based dynamic quantizer defined in \eqref{eq: quantizer-1}-\eqref{eq: quantizer-3} on both sides of encoder and decoder, this paper first assumes that there exists a noiseless feedback channel that reliably delivers acknowledge signals from the decoder to the encoder to indicate a successful reception of a block of bits. With the noiseless feedback assumption, $\hat{\overline{x}}(t_{k}^{+})$ and $U(t_{k}^{+})$ can be implemented synchronously on both sides of encoder and decoder by sharing $R_{k}$. Besides feasible implementation, the dynamic quantizer must guarantee that the hypercubic box defined by $\hat{\overline{x}}(t_{k}^{+})$ and $U(t_{k}^{+})$ contains vehicle state $x$ for any time instant $t_{k}, k \in \mathbb{Z}_{\geq 0}$. The method to construct functions $g_{\overline{x}}$ and $g_{U}$ is discussed in Section \ref{subsec: encoder-decoder}.

\begin{remark}
	The event-based dynamic quantization method differs from existing frameworks
	\cite{nesic2009unified,liberzon2005stabilization} in two aspects. Firstly, unlike the constant quantization level~(constant data rate) considered in prior work \cite{nesic2009unified,liberzon2005stabilization}, the quantization level~($2^{nR_{k}}$) in the event-based dynamic quantizer is time varying and stochastic changes as a function of the vehicle state. Secondly, the transmission time instants are generated sporadically rather than scheduled with an equidistant time interval. These differences distinguish our results from others. 
\end{remark}

\section{Problem Formulation}
\label{section: problem-formulation}
Let $\overline{x}(t)=x(t)-x^{d}$ denote the tracking error and $e(t) \triangleq \overline{x}(t)-(\hat{x}(t)-x^{d})=x(t)-\hat{x}(t)$ denote the estimation error induced by the bursty fading channel. The closed-loop dynamics of the VNS defined in \eqref{eq:sys}, \eqref{ineq:EBB}, \eqref{eq:controller} and \eqref{eq: quantizer-1}-\eqref{eq: quantizer-3} can be reformulated as a stochastic hybrid system defined as below,
\begin{subequations}
	\label{shs}
\begin{align}
\dot{\overline{x}}(t)&=f_{\overline{x}}(\overline{x}, e, w), \quad \forall t \in (t_{k}, t_{k+1})  \label{shs: x}\\
\dot{e}(t)&=f_{e}(\overline{x}, e, w), \quad \forall t \in (t_{k}, t_{k+1}) \label{shs: e}\\
U(t_{k+1})&=f_{U}(T_{k+1}, R_{k},\overline{x}(t_k), e(t_{k}^{+}), U(t_k)) \label{shs: U}\\
e(t_{k}^{+})&=h_{e}(T_{k}, R_{k}, \overline{x}(t_k), e(t_k), U(t_k))  \label{shs: jump-e} \\
U(t_{k}^{+})&=h_{U}(R_{k}, U(t_{k})) \label{shs: jump-U}
\end{align}
\end{subequations}
where 
\begin{align*}
f_{\overline{x}}(\overline{x}, e, w)& \triangleq f(\overline{x}+x^{d}, \kappa(\overline{x}-e), w) \\
f_{e}(\overline{x}, e, w)& \triangleq f_{\overline{x}}(\overline{x}, e, w)-f(\overline{x}-e+x^{d}, \kappa(\overline{x}-e), 0) \\
f_{U}(T, R,\overline{x}, e^{+}, U)& \triangleq g_{U}(T, \overline{x}-e^{+}+x^{d}, 2^{-R}U) \\
h_{e}(T, R, \overline{x}, e, U) & \triangleq \overline{x}-g_{\overline{x}}(T, R, \overline{x}-e+x^{d}, U)\\
h_{U}(R, U) &\triangleq 2^{-R}U.
\end{align*}
Equations \eqref{shs: x} and \eqref{shs: e} represent the continuous dynamics in the stochastic hybrid framework, while equation \eqref{shs: U} characterizes a controlled discrete time stochastic process. Equations \eqref{shs: jump-e} and \eqref{shs: jump-U} represent the stochastic jumps for the continuous and discrete states, respectively. The randomness of this stochastic hybrid systems comes from the stochastic process $\{R_{k}\}_{k=0}^{\infty}$ that is assumed to satisfy the SD-EBB in \eqref{ineq:EBB}. 

Under the stochastic hybrid framework, the objective of this paper is to design an event based communication scheme to ensure \emph{stochastic stability} for VNSs in \eqref{shs}. In particular, this paper considers both sample-based and mean stability. Sample-based stability emphasizes the behavior of almost all sample paths toward or around the origin while mean stability stresses system behavior in expectation. The formal definitions are provided as below,
\begin{definition}[Stochastic Stability \cite{kozin1969survey}]
	\label{def:ss}
Consider a stochastic hybrid framework defined in \eqref{shs}, and let $\overline{x}_0 \triangleq x(0)-x^{d} \in \mathbb{R}^{n}$ denote the initial state,
	\begin{itemize}
		\item[\textbf{E1}] The system in \eqref{shs} with $w=0$ is \emph{asymptotically stable in expectation} with respect to origin, if for any given $\epsilon > 0$, there exists $\delta(\epsilon)$ such that $|\overline{x}_0| \leq \delta$ implies 
		\begin{align}
		\mathbb{E}\big\{|\overline{x}(t)| \big\} < \epsilon 
		\end{align}
		and $\lim_{t \rightarrow \infty}\mathbb{E}\big\{|\overline{x}(t)| \big \}=0$.
		\item[\textbf{E2}] The system in \eqref{shs} with $|w|_{\mathcal{L}_{\infty}} \leq M$ is \emph{uniformly asymptotically bounded in expectation}, if for a given $(\Delta(M), \Delta_0(M))$ with $\Delta_0, \Delta > 0$, there exists a $\epsilon (M, \Delta_0) > 0 $ such that for $|\overline{x}(0)| \leq \Delta_0$,
		\begin{align}
	\mathbb{E}\big\{|\overline{x}(t)| \big\} \leq \epsilon(M, \Delta_0) \quad \forall t \in \mathbb{R}_{\geq 0}
		\end{align}
		and $\lim_{t \rightarrow \infty}\mathbb{E}\big\{|\overline{x}(t)| \big\} \leq \Delta(M)$.
		\item[\textbf{P1}] The system in \eqref{shs} with $w=0$ is \emph{almost surely asymptotically stable} with respect to origin, if for any given $\epsilon, \epsilon' > 0$, there exists $\delta(\epsilon,\epsilon')$ such that $|\overline{x}_0| \leq \delta$ implies 
		\begin{align}
		\label{ineq:as-as}
		{\rm Pr}\big\{\sup_{t \geq 0} |\overline{x}(t)| \geq \epsilon' \big\} < \epsilon 
		\end{align}
		and $\lim_{\tau \rightarrow \infty}{\rm Pr}\big\{\sup_{t \geq \tau} \big |\overline{x}(t)|\geq \epsilon'\big\}=0$.
		\item[\textbf{P2}] The system in \eqref{shs}  with $|w|_{\mathcal{L}_{\infty}} \leq W$ is \emph{practically stable in probability} if for a given $(\Delta(M), \Delta_0(M))$ with $0 < \Delta_0 < \Delta$ and for any $\epsilon' > 0$, there exists a $\epsilon (M, \Delta) > 0 $ such that for $|\overline{x}(0)| \leq \Delta_0$,
			\begin{align}
			\lim_{t \rightarrow +\infty}{\rm Pr}\{|\overline{x}(t)| \geq \Delta + \epsilon' \} \leq \epsilon(M, \Delta).
			\end{align}
	\end{itemize}
\end{definition}
\begin{remark}
	Among all four definitions of stochastic stability, the \emph{almost sure asymptotic stability} is the strongest notion of stability, which requires almost all the samples of the system in \eqref{shs} asymptotically converge to origin with probability one. The notion of \emph{mean stability}~(\textbf{E1}) is weaker in the sense that it only requires the expected value of the system trajectory's magnitude asymptotically to go to zero. In general, \emph{mean stability} does not imply \emph{almost sure asymptotic stability} while the later certainly implies the former. For more discussions on stochastic stability, please refer to \cite{kozin1969survey,Kushner1967}.
\end{remark}
\begin{remark}
 When a non-varnishing but bounded external disturbance is present in VNS, the asymptotic stability~(i.e., \textbf{E1} and \textbf{P1}) cannot be guaranteed.  Relaxed stability notions defined in \textbf{E2} and \textbf{P2} are therefore introduced to characterize the system behavior  around a compact set in expectation or in probability. Specifically, the \emph{uniformly asymptotic boundedness in expectation}~(\textbf{E2}) requires that the expectation of the norm of system states is uniformly bounded and asymptotically converges to a constant that depends on the magnitude of external disturbance. The notion of   \emph{practical stability in probability} requires that the probability~(\textbf{P2}) of the system trajectories leaving a compact set is bounded from above by a function that depends on both the magnitude of the external disturbance and the size of the compact set. By Markov's Inequality, it is straightforward to show that the notion \textbf{E2} implies \textbf{P2}.
\end{remark}
\section{Assumptions}
\label{sec:assumption}
This section presents two main assumptions that are needed to establish our main results. The first assumption is the \emph{input to state stability} for the subsystem $\overline{x}$ defined in \eqref{shs: x} and is stated formally as below. 
\begin{assumption}
\label{assumption: x}
Consider the stochastic hybrid system in \eqref{shs}, the subsystem $\overline{x}$ defined in \eqref{shs: x} is \emph{input-to-state stable}~(ISS) from $\overline{x}$ to the estimation error $e$ and external disturbance $w$. In particular, we assume that there exist a concave class $\mathcal{KL}$ function $\beta(\cdot, \cdot)$,  a class $\mathcal{K}$ function $\chi_{2}(\cdot)$ and a linear function $\overline{\chi}_{1} > 0$ such that
\begin{align}
|\overline{x}(t)| \leq \beta(|\overline{x}(t_{0})|, t-t_{0}) + \overline{\chi}_{1}(|e|_{[t_{0}, t]}) + \chi_{2}(|w|_{[t_{0}, t]})
\end{align}
The subsystem $\overline{x}$ is \emph{exponentially input-to-state stable}~(Exp-ISS) if $\beta(\cdot, \cdot)$ is an exp-$\mathcal{KL}$ and $\chi_{2}(\cdot)$ is a linear function. 
\end{assumption}
\begin{remark}
The ISS assumption is used to ensure \emph{stability in expectation}~(\textbf{E1} and \textbf{E2} in Definition \ref{def:ss}) for the system in \eqref{shs} while the assumption of exp-ISS is needed to ensure almost sure asymptotic stability~(\textbf{P1} in Definition \ref{def:ss}) .  
\end{remark}
\begin{assumption}
\label{assumption: e}
For given $0 < w_{1} < w_{2}$ and $L_{x}, L_{e}, L_{w} \in \mathbb{R}_{+}$, suppose there exists a Lyapunov function $W(e)$ such that the subsystem $e$ in Equation (\ref{shs: e}) satisfies
\begin{subequations}
\begin{align}
w_{1}|e| &\leq W(e) \leq w_{2}|e| \\
\nabla W(e) f_{e}(\overline{x}, e, w) &\leq L_{e} W(e) + L_{x}|\overline{x}|+ L_{w} |w|
\label{ineq: W(e)}
\end{align} 
\end{subequations}
\end{assumption}
\begin{remark}
The Assumption \ref{assumption: e} is equivalent to the assumption that the subsystem $e$ is Exp-ISS with respect to $x$ and $w$ \cite{nevsic2004input}. This assumption is weaker than the uniformly Lipschitz assumption in \cite{hu2014event} in the sense that the former one includes the latter one as a special case ($L_{x}=0$). To see this, the uniformly Lipschitz assumption suggests that there exists a $L_{f} > 0$ such that
\begin{align*}
|f(x, u, w)-f(\hat{x}, u, 0)| \leq L_{f}(|x-\hat{x}|+|w|), \forall, x, \hat{x}\in \Omega_{x}
\end{align*}
where $\Omega_{x}$ is a compact set. This uniformly Lipschitz assumption on the vector field $f$ implies that 
\begin{align*}
\frac{d |e|}{dt} \leq L_{f}|e|+L_{f}|w|
\end{align*}
which is a special case of \eqref{ineq: W(e)} with $L_{x}=0$.
\end{remark}

\section{Main Results}
\label{sec: main-results}
This section presents the main results of developing self-triggered communication scheme to ensure \emph{stochastic stability} for VNS defined in \eqref{shs}. In particular, self-triggered transmission schemes are designed to generate sporadic transmission sequence $\{t_k\}_{k=0}^{\infty}$ under which the VNS in \eqref{shs} is either \emph{asymptotically stable in expectation}~(Theorem \ref{theorem:as-in-expectation}) or \emph{almost surely asymptotically stable} (Theorem \ref{theorem:almost-sure-as}) without external disturbance~($w=0$), and either \emph{uniformly asymptotically bounded in expectation}~(Theorem \ref{thm: ua-bounded}) or \emph{practically stable in probability} (Theorem \ref{theorem:pc}). Under the self-triggered transmission scheme, another result~(Proposition \ref{prop: encoder-decoder} ) of this paper is to construct a feasible event-based encoder-decoder pair~(i.e., function $g_x$ in \eqref{eq: quantizer-1} and $g_{U}$ in \eqref{eq: quantizer-3}) in which the event-based dynamic quantizer do not saturate at any transmission time instant, that is, the system state in \eqref{shs} is guaranteed to be captured by the proposed event-based encoder-decoder scheme. 

Before stating the main theorems, one needs the following technical lemma to show that the expectation of the quantization resolution $\mathbb{E}(2^{-R_k})$ can be bounded by a function of the vehicle state when a state-dependent bursty fading channel is present.
\begin{lemma}
	\label{lem:G}
	Consider the EBB characterization in (\ref{ineq:EBB}), define a function $G(s)=e^{-h(s)\gamma(s)}(1+h(s)\gamma(s)), s \in \mathbb{R}_{+}$, then
	\begin{align}
	\label{ineq:q_level}
	\mathbb{E}(2^{-R_{k+1}}) \leq G(|x(k+1)|)
	\end{align}
	and $G(s) \in [0, 1], \forall s \in \mathbb{R}_{+}$ is a strictly increasing function with
	\begin{align}
	G(s) \rightarrow 0 &\iff h(s)\gamma(s) \rightarrow +\infty \\  G(s) \rightarrow 1 &\iff h(s)\gamma(s) \rightarrow 0
	\end{align}
\end{lemma}
\begin{IEEEproof}
	By EBB model in (\ref{ineq:EBB}), equivalently
	\begin{align*}
	{\rm Pr}\{2^{-R_{k+1}} \geq -h(|x(k)|)+\sigma\} \leq e^{-\gamma(|x(k)|)\sigma}
	\end{align*}
	Since
	$E[2^{-R_{k+1}}]=\int_{0}^{\infty}{\rm Pr}\{2^{-R_{k+1}} \geq y\}dy$, 
	following the same argument in \cite{bin2013}, one obtains $\mathbb{E}(2^{-R_{k+1}}) \leq G(|x(k+1)|)$. The details of the proof are omitted here for the limitation of space. Taking the first derivative of function $G(s)$ with respect to $s$ yields
	\begin{align*}
	\frac{dG(s)}{ds} =-\frac{d(h(s)\cdot\gamma(s))}{ds} h(s)\gamma(s)e^{-h(s)\gamma(s)} \geq 0
	\end{align*}
	and $\frac{dG(s)}{ds}=0 \iff h(s)\gamma(s)=0$ because $h(s)$ and $\gamma(s)$ are nonnegative strictly decreasing functions. Therefore, $G(s)$ is a strictly increasing function. Let $y:=h(s)\gamma(s) \in \mathbb{R}_{+}$, then it is easy to check that $G(y)=e^{-y}(1+y)$ is strictly decreasing with respect to $y$. Since $y \geq 0$, we have $\lim_{y \rightarrow 0}G(y) \rightarrow 1$ and $\lim_{y \rightarrow +\infty}G(y) \rightarrow 0$.
\end{IEEEproof}
\begin{remark}
	The function $G(\cdot)$ is directly related to the functions $h(\cdot)$ and $\gamma(\cdot)$ in the SD-EBB characterization in \eqref{ineq:EBB} and can therefore be viewed as a priori knowledge of the state-dependent fading channel.
\end{remark}

Inequality \eqref{ineq:q_level} implies that quantization error decreases as the vehicle state $x$ approaches its origin.  It is easy to see that $G(|x|) \rightarrow 1$ as $h(|x|) \rightarrow 0$,  which corresponds to the scenario where the vehicles are far apart and beyond communication range. This paper will focus on the situation when the vehicles are within communication range and therefore the SD-EBB model provides a reasonable bound on the channel conditions. In particular, let $\Omega_{c}=\{x \in \mathbb{R}^{n} \big\vert |x| \leq M_c \} $ with $M_c > 0$ denoting the region that communications between vehicles are available, and $G(|x|) < 1, \forall x \in \Omega_{c}$. From communication's standpoint, the communication range $M_c$ can be enlarged by increasing the transmission power. 

Since $G(\cdot) \in [0, 1]$ is a continuous and strictly monotonically increasing function, the inverse of $G(\cdot)$ exists and is also continuous, strictly monotonically increasing. Thus,  let $\Omega_{x}=\{x \in \mathbb{R}^{n} \big\vert |x|<G^{-1}( w_{1}/w_{2})\}$ denote a region of interest for vehicle states with $w_1, w_{2} > 0$ defined in Assumption \ref{assumption: e}, and it is straightforward to show that $\Omega_{x}$ is a nonempty and compact set. Furthermore, one has $\Omega_{x} \subset \Omega_{c}$ if $w_{1} < w_{2}$.  The stability results in Section \ref{subsec: stability} will be examined under the situation that vehicles are within the communication range, i.e.,   $x \in \Omega_{x} \subset \Omega_{c}$.
\subsection{Self-triggering to Achieve Stochastic Stability}
\label{subsec: stability}
A self-triggered scheme is developed in this section to ensure the notions of stochastic stability defined in Definition \ref{def:ss}. With Assumptions \ref{assumption: x} and \ref{assumption: e},  this section first presents two theorems showing that the VNS defined in \eqref{shs} can asymptotically track pre-defined set-points in expectation~(Theorem \ref{theorem:as-in-expectation}) under the ISS assumption or almost surely~(Theorem \ref{theorem:almost-sure-as}) under the exp-ISS assumption if external disturbance $w=0$ is absent. 
\begin{theorem}
\label{theorem:as-in-expectation}
Consider the stochastic hybrid system in \eqref{shs} without external disturbance ($w=0$) and the EBB channel model in (\ref{ineq:EBB}),  suppose the ISS assumption in Assumptions \ref{assumption: x} and Assumption \ref{assumption: e} hold, the system is asymptotically stable in expectation with respect to the origin, if the transmission time instants $\{t_{k}\}$ are generated by  
\begin{align}
\label{TI}
t_{k+1}=t_{k}+\frac{1}{L_{e}}\ln \Big(1+ \frac{1-\frac{w_{2}}{w_{1}}G( |x(t_{k})|)}{\frac{w_{2}}{w_{1}}G(|x(t_{k})|)+\frac{L_{x}\overline{\chi}_{1}}{L_{e}w_{1}}} \Big).
\end{align}
Furthermore, if the vehicle system is within the communication range, i.e., $x \in \Omega_{x}$, there exists $\underline{\Delta}> 0$ such that the transmission time interval $t_{k+1}-t_{k} \geq \underline{\Delta}, \forall k \in \mathbb{Z}_{+}$, i.e. the self triggered scheme in \eqref{TI} assures Zeno-free behavior.
\end{theorem}
\begin{IEEEproof}
	See Appendix \ref{appendix}.
\end{IEEEproof}
\begin{remark}
	The transmission time intervals $\{T_{k}\}_{k=0}^{\infty}$ with $T_{k}=t_{k+1}-t_{k}$ generated by \eqref{TI} monotonically increases when vehicle states $|x(t)|$, such as inter-vehicle distance and bearing angles, move towards the origin. This property  implies that the VNS can transmit less frequently when a good channel condition is guaranteed by either a short inter-vehicle distance or aligned directional antennae mounted in both vehicles. The function $G$ in \eqref{TI} that is derived from the proposed SD-EBB model in \eqref{ineq:EBB} quantitatively assesses how channel conditions vary as a function of vehicle states. Such quantitative measurement is then used to construct a self triggered scheme that adapts its transmission frequency by taking into account the dynamic interactions between physical vehicle systems and communication channels.  
\end{remark}
\begin{theorem}
\label{theorem:almost-sure-as}
Suppose all the conditions and assumptions in Theorem \ref{theorem:as-in-expectation} hold, and suppose the subsystem $\overline{x}$ is Exp-ISS~(i.e., Exp-ISS in Assumption \ref{assumption: x}) holds,  the VNS in \eqref{shs} without external disturbance is \emph{almost surely asymptotically stable} with respect to origin if the transmission time sequence is recursively generated by \eqref{TI}.  The non-Zeno transmission is guaranteed if vehicles are within the communication range (i.e., $x \in \Omega_{x}$).
\end{theorem}
\begin{IEEEproof}
The proof is provided in Appendix \ref{appendix}.
\end{IEEEproof}

Since the strong notion of \emph{asymptotic stability} can not be guaranteed in the presence of non-varnishing disturbance, this section shows that weak notions of \emph{uniformly asymptotic boundedness in expectation}~(\textbf{E2}) and \emph{practical stability in probability}~(\textbf{P2}) can be achieved under the self triggered scheme defined in \eqref{TI}. 

\begin{theorem}
	\label{thm: ua-bounded}
Consider the VNS defined in \eqref{shs} with essentially bounded external disturbance $|w|_{\mathcal{L}_{\infty}} \leq M$, and suppose the fading channel satisfies the SD-EBB characterization defined in \eqref{ineq:EBB}.  Suppose the ISS assumption in Assumption \ref{assumption: x} and Assumption \ref{assumption: e} hold, if the transmission time sequence $\{t_{k}\}$ is generated by \eqref{TI}, then the system in \eqref{shs} is uniformly asymptotically bounded in expectation~(\textbf{E2}).	
\end{theorem}
\begin{IEEEproof}
	See Appendix \ref{appendix}.
\end{IEEEproof}

\begin{theorem}
\label{theorem:pc}
Suppose the hypothesis in Theorem \ref{thm: ua-bounded} holds, then the system in \eqref{shs} is practically stable in probability~(\textbf{P2}). More specifically, there exists a class $\mathcal{KL}$ function $\beta_{\epsilon}(\cdot, \cdot)$ such that 
\begin{align}
\label{ineq: ps-in-probability}
\lim_{t \rightarrow +\infty}{\rm Pr}\{|\overline{x}(t)| \geq \Delta + \epsilon \} \leq \beta_{\epsilon}(M, \Delta).
\end{align}
\end{theorem}
\begin{IEEEproof}
	See Appendix \ref{appendix}.
\end{IEEEproof}

\begin{remark}
The probability bound in \eqref{ineq: ps-in-probability} measures the safety level as a function of the size of a safe region $\Delta$ as well as the magnitude of external disturbance $M$. This safety metric provides a trade-off between the choices of $\Delta$ and $M$, which shows that the system is more likely to be safe with a smaller magnitude of external disturbance $M$ and a larger safety region $\Delta$. 
\end{remark}
\subsection{Design of Event-based Encoder and Decoder}
\label{subsec: encoder-decoder}
The stability results hold under the hypothesis that the vehicle state $\overline{x}(t_k)$ at each time instant $t_{k}, \forall k \in \mathbb{Z}_{\geq 0}$ must be captured by the encoder and decoder defined in \eqref{eq: quantizer-1}-\eqref{eq: quantizer-3} with parameters $(\hat{\overline{x}}(t_k^{+}), U(t_{k}^{+}))$ representing the centroid and size of the quantizer respectively. This hypothesis is proved in the following proposition by showing that an \emph{Encoder/Decoder} pair can be designed to recursively construct and synchronize the parameters $(\hat{\overline{x}}(t_k^{+}), U(t_{k}^{+}))$ as time increases. For notational simplicity, let $\hat{\overline{x}}_{k}^{+}:=\hat{\overline{x}}(t_{k^{+}})$ and $U_{k}^{+}:=U(t_{k}^{+})$.
\begin{proposition}
	\label{prop: encoder-decoder}
	Suppose Assumptions \ref{assumption: x} and \ref{assumption: e} hold, and let $\{t_k\}_{k=0}^{\infty}$ denote the transmission time sequence generated by \eqref{TI} and $T_k=t_{k+1}-t_k$. Suppose the initial information pair $(\hat{\overline{x}}_{0}^{+}, U_{0}^{+}))$ and the number of successfully received bits $R_k, \forall k \in \mathbb{Z}_{\geq 0}$, are known to the \emph{Encoder} and \emph{Decoder} by noiseless feedback channels, if the sequence of information pairs $\{\hat{\overline{x}}_{k}^{+}, U_{k}^{+}\}_{k=1}^{\infty}$ is constructed by 
	\begin{subequations}
		\label{eq: information-pair}
		\begin{align}
		\label{eq: U_k}
		&U_{k+1}^{+}=\frac{2^{-R_{k+1}}}{\eta_{k+1}}\bigg(\frac{v_{2}}{v_{1}}e^{L_{e}T_{k}}U_k^{+}+\frac{e^{L_{e}T_k}-1}{v_{1}L_{e}}\Big(L_{x}\beta(|\hat{\overline{x}}_{k}^{+}|+U_k^{+}, 0) \nonumber\\
		& +L_{w}W+L_{x}\alpha_{2}(W)\Big)\bigg),\\
		&\hat{\overline{x}}_{k+1}^{+}=U_{k+1}^{+}2^{R_{k+1}}\sum_{j=1}^{R_{k+1}}\frac{1}{2^j}q\big(b_{j}(k+1)\big)
		+\Phi(\hat{\overline{x}}_{k}^{+},T_{k}),
		\label{eq: x_q}
		\end{align}
	\end{subequations}
	where
	\begin{align*}
	\eta_{k}=1-\frac{L_x\overline{\alpha}_{1}}{w_{1}L_e}(e^{L_{e}T_{k}}-1) > 0,
	\end{align*}
	and $\Phi(s,t)$ is the solution to the following differential equation
	\begin{align}
	\label{eq:ode}
	\dot{\overline{x}}=f_{\overline{x}}(\overline{x}, 0, 0), \quad  \overline{x}(0)=s.
	\end{align}
	where $f_{\overline{x}}(\cdot, \cdot, \cdot)$ is defined in \eqref{shs: x} and $q(\cdot): \{0, 1\}^{n} \rightarrow \{-1, 1\}^{n}$ is a function that maps the binary value of the bit vector to a $n$ dimensional vector whose elements are $\pm 1$, i.e., 
	 \begin{align*}
	 q_{i}(b)=\begin{cases}
	 1 &  \text{if the $i^{th}$ bit in bit-vector $b$ is $1$}, \\
	 -1 & \text{otherwise}.
	 \end{cases}
	 \end{align*}
	then the estimation error $e(k)=\overline{x}(t_k)-\hat{\overline{x}}_{k}^{+}$ is bounded as
	\begin{align}
	|\overline{x}(t_k)-\hat{\overline{x}}_{k}^{+}| \leq U_k
	\end{align}
	for all $k \in \mathbb{Z}_{\geq 0}$.
\end{proposition}
\begin{IEEEproof}
	See Appendix \ref{appendix}.
\end{IEEEproof}
\begin{remark}
$\eta_{k} > 0$ holds $\forall k \in \mathbb{Z}_{+}$ if the self-triggered scheme in \eqref{TI} is adopted. The recursive functions in \eqref{eq: U_k} and \eqref{eq: x_q} correspond to the Encoder/Decoder structure defined in \eqref{eq: quantizer-1}-\eqref{eq: quantizer-3}.  The Encoder/Decoder design in \eqref{eq: U_k} and \eqref{eq: x_q} generalizes the result in \cite{hu2014event}. One can recover the Encoder/Decoder structure in \cite{hu2014event} by setting $L_{x}=0$. This generalization is possible due to Assumption \ref{assumption: e} which is weaker than the uniformly Lipschitz assumption in \cite{hu2014event}.
\end{remark}
\begin{remark}
Equation \eqref{eq: x_q} is a recursive rule updating the centroid of a dynamical uniform quantizer \cite{martins2006feedback}. The structure of the solution $\Phi(s, t)$ can be determined offline by solving the nonlinear differential equation \eqref{eq:ode}~(nominal system without considering the network effect) with an initial value. In general, obtaining an analytic solution for a nonlinear ODE \eqref{eq:ode} is difficult, but one can obtain approximation on the solution by integrating the function $f_{\overline{x}}$ from $t_{k}$ to $t_{k}+T_k$, i.e., $\Phi(\hat{\overline{x}}_{k}^{+}, T_k)=\hat{\overline{x}}(t_{k}+T_{k})=\hat{\overline{x}}(t_k^{+})+ \int_{t_{k}}^{t_{k}+T_{k}}f_{\overline{x}}(\hat{\overline{x}}(t), 0, 0)dt$. However, the analytic solution can be obtained if the function $f_{\overline{x}}$ is linear, e.g., $f(\overline{x}, 0, 0)=A\overline{x}$, then one has $\Phi(\hat{x}_{k}^{+},T_{k})=\exp(AT_{k})\hat{x}_{k}^{+}$.
\end{remark}
\section{Leader-Follower $\alpha-L$ Formation Control}
\label{sec: leader-follower}
\begin{figure}[!ht]
	\centering
	\includegraphics[width=0.4\textwidth]{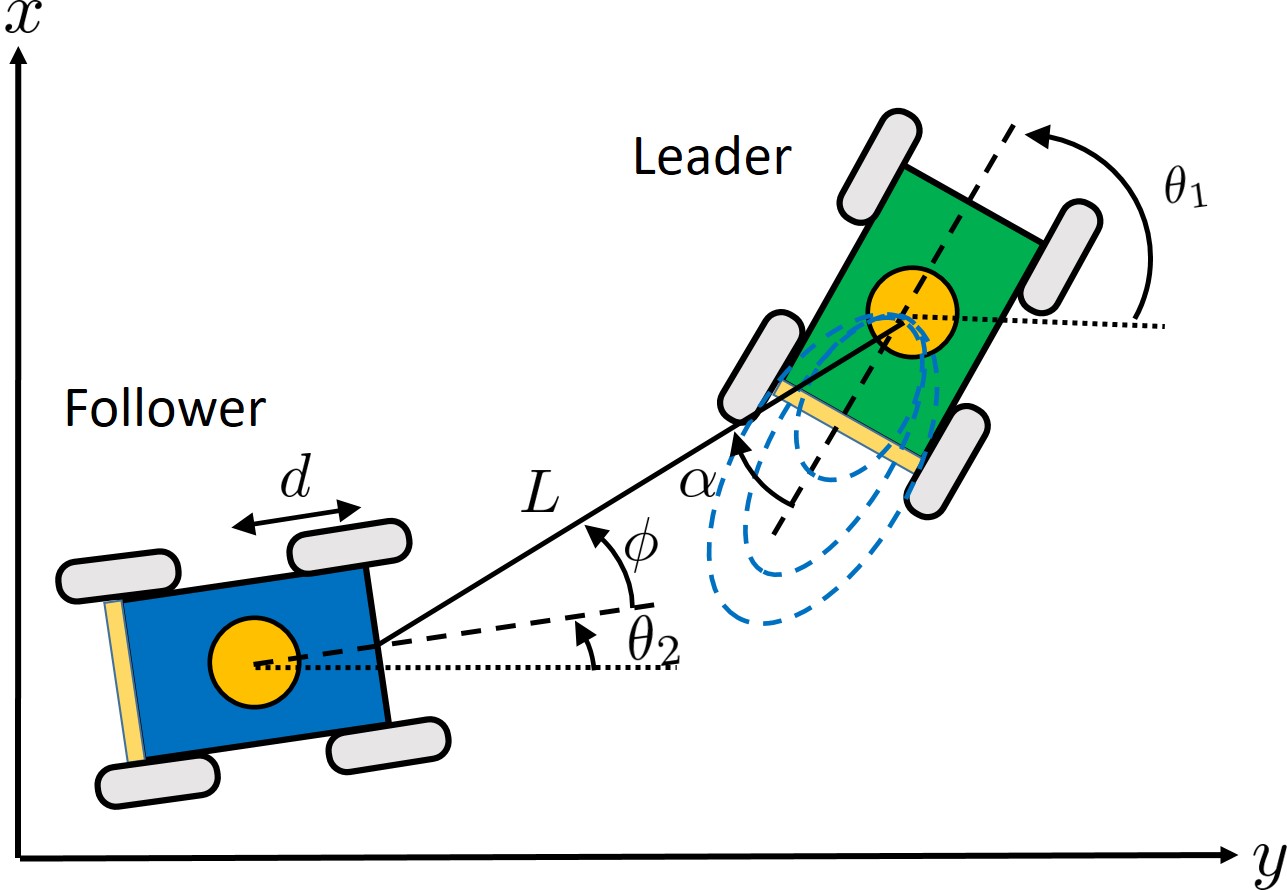}
	\caption{Leader Follower $\alpha-L$ Formation Control}
	\label{fig:LF}
\end{figure}
The main results in Section \ref{sec: main-results} can be illustrated via a leader-follower $\alpha-L$ control example shown in Fig. \ref{fig:LF}. Both vehicles satisfy the following kinematic model:
\begin{align}
\label{eq:km}
\dot{p}_{x, i}=v_{i}\cos{\theta_{i}}, \quad  \dot{p}_{y, i}=v_{i}\sin{\theta_{i}}, \quad \dot{\theta_{i}}=\omega_{i}.
\end{align}
where $p_{x, i}, p_{y, i}, i=1,2 $ are the horizontal and vertical positions of leader~($i=1$) and follower~($i=2$), respectively, and $\theta_{i}, i=1, 2$ are orientations of leader and follower relative to the horizon.  Based on the kinematic model in \eqref{eq:km}, 
the leader-follower $\alpha-L$ system in Fig. \ref{fig:LF} satisfies the following ODE \cite{bin2014-2},
\begin{align}
	\label{eq:alpha-L}
	\begin{array}{lcl}
		\dot{L} &=& v_{1} \cos \alpha - v_{2} \cos \phi - d \omega_{2} \sin \phi \\
		\dot{\alpha} &=& \frac{1}{L} \left( -v_{1} \sin \alpha - v_2 \sin \phi + d \omega_2 \cos \phi \right) + \omega_{1}
	\end{array}
\end{align}
where $d$ is the length from the center of the vehicle to its front. $v_{1}$ and $\omega_{1}$ are leader's speed and angular velocity, while $v_{2}$ and $\omega_{2}$ are follower's speed and angular velocity. $L$ is the inter-vehicle distance that is measurable by both leader and follower, $\alpha$ and $\phi$ are relative bearing angles of leader to follower and follower to leader respectively. It is assumed that $\alpha$ is only measurable to the leader, and $\phi$ is available for the follower. What is not directly known to the follower is the bearing angle $\alpha$. Therefore, the leader-follower pair characterizes a system taking the form in Equation \eqref{shs: x} that requires the leader to transmit its bearing angle $\alpha$ to the follower over a wireless communication channel. The wireless channel is accessed by a directional antenna that is mounted at the back of the leader where the channel exhibits exponential burstiness and satisfies the state dependent EBB characterization in (\ref{ineq:EBB}) with $(L, \alpha)$ as the vehicle state $x$. As shown in Fig. \ref{fig:LF}, the directional antenna has a radiation range from $-\frac{\pi}{2} \leq \alpha \leq \frac{\pi}{2}$ out of which the communication channel is assumed to be zero.  

With limited information on $\alpha$, the control objective is to have the follower adjust its speed $v_{2}$ and angular velocity $\omega_{2}$ to achieve desired inter-vehicle separation $L_{d}$ and bearing angle $\alpha_{d}$ almost surely in the presence of deep fades. A standard input to state feedback linearization method is used to generate control inputs $(v_{2}, \omega_{2})$ over each transmission time interval $[t_k, t_{k+1})$,
\begin{align}
	\left [ \begin{array}{c} v_{2} \\ \omega_{2} \end{array} \right]&=\left [ \begin{array}{cr} -\cos{\phi} & -L\sin{\phi} \\ -\frac{\sin{\phi}}{d} & \frac{L}{d}\cos{\phi} \end{array} \right] \left ( \left [ \begin{array}{c} K_{L}(L_{d}-L) \\ K_{\alpha}(\alpha_{d}-\hat{\alpha}) \end{array} \right] \right. \nonumber \\
	&\left. -\left [\begin{array}{cr} \cos{\hat{\alpha}} & 0 \\ -\frac{\sin{\hat{\alpha}}}{L} & 1  \end{array}\right ] \left [ \begin{array}{c} v_1 \\ \omega_{1} \end{array} \right ] \right)
	\label{eq:control}
\end{align}
where $(K_{L}, K_{\alpha})$ are the controller gains. $\hat{\alpha}$ represents the prediction of the bearing angle over $[t_k, t_{k+1})$ that satisfies
\begin{align}
	\dot{\hat{\alpha}}=K_{\alpha}(\alpha_{d}-\hat{\alpha}), \hat{\alpha}(t_k)=\alpha^{q}({t_k})
	\label{eq:pre_alpha}
\end{align}
with the estimate of the bearing angle $\alpha^{q}({t_k})$ as the initial value.

Assume that the leader changes its speed $v_{1}=g_{v}(L)+n_1$ and angular velocity $\omega_{1}=g_{\omega}(\alpha)+n_2$ as a function of the inter-vehicle distance $L$ and its relative bearing angle $\alpha$ with essentially bounded disturbance $|n_i|_{\infty} \leq M, i=1,2$. With the controller in (\ref{eq:control}) and (\ref{eq:pre_alpha}), the inter-vehicle distance $L$ and the bearing angle $\alpha$ therefore satisfies the following differential equations over time interval $[t_k, t_{k+1})$. 
\begin{align}
	\label{eq:closed-loop}
	\begin{array}{lcl}
		\dot{L}&=&K_{L}(L_d-L)+(g_{v}(L)+n_1)(\cos{\alpha}-\cos{\hat{\alpha}}) \\
		\dot{\alpha}&=&\frac{(g_{v}(L)+w_1)}{L}(\sin{\hat{\alpha}}-\sin{\alpha})+K_{\alpha}(\alpha_d-\hat{\alpha})\\
		&&+ g_{\omega}(\alpha)+n_2-g_{\omega}(\hat{\alpha})
	\end{array}
\end{align}
for all $k \in \mathbb{Z}_{+}$. One can easily see that the closed loop system in (\ref{eq:closed-loop}) represents one of the structures in (\ref{eq:sys}) and (\ref{eq:controller}). The results in Section \ref{sec: main-results} can be directly applied to leader-follower $\alpha-L$ formation control problem. 
\section{Simulation and Experimental Results}
\label{sec: simulation}
This section first presents simulation results examining the advantages of the proposed self triggered scheme over traditional event triggered scheme in \cite{wang2011attentively} in the leader-follower example. These simulation results are further verified in the MobileSim Robot Simulator with real parameters of Pioneer 3 DX mobile robots. The results generated by MobileSim are proved to provide comparatively close performance to real experiments \cite{Mobilesim}. 

A two-state Markov chain model is used to simulate the
fading channel between the leader and the follower. The two-state Markov chain model has one state
representing the good channel state and the other representing
the bad channel state. The good channel state means that
the transmitted bit is successfully received, while the bad
state means that the transmitted bit is lost. Following the
two-state Markov chain model in \cite{zhang1999finite}, this simulation uses $p_{12}=0.08\sqrt{\frac{\pi}{2}r}$ to represent the transition probability from good state to bad state, and $p_{21}=0.08\sqrt{\frac{\pi}{2}}\frac{\sqrt{r}}{e^{0.25r}-1}$ to represent the transition probability from bad state to good state, where $r=\frac{L}{p \cos{\alpha}}$ and $p$ is the transmission power. It is clear that the transition matrix for this two-state Markov chain model is a function of the vehicular states ($L$ and $\alpha$) for a fixed transmission power $p$. Following the results in \cite{hu2015distributed}, the SD-EBB functions used in this simulation are $h(\alpha, L)=0.8\overline{R}e^{-0.25\frac{L}{p \cos{\alpha}}}$ and $\gamma(\alpha, L)=8\frac{p \cos{\alpha}}{L}$, with $\overline{R}=4$ as the total number of bits transmitted over each time interval and $p=8$ as the transmission power level. The initial inter-vehicle distance and bearing angle are $L(0)=15~m$ and $\alpha(0)=-30^{\circ}$. The controller gains are $K_{L}=K_{\alpha}=1$. Let the leader's speed $v_1$ and angular velocity $\omega_1$ be $v_1=0.8 L$ and $\omega_1=2.2 \alpha$, respectively. The theoretical results are verified based on Monte Carlo simulation method under which each simulation example is run $100$ times over a time interval from $0$ to $10$ second.
\subsection{Simulation Results in Matlab}
\begin{figure}[!ht]
	\centering
	\includegraphics[width=0.5\textwidth]{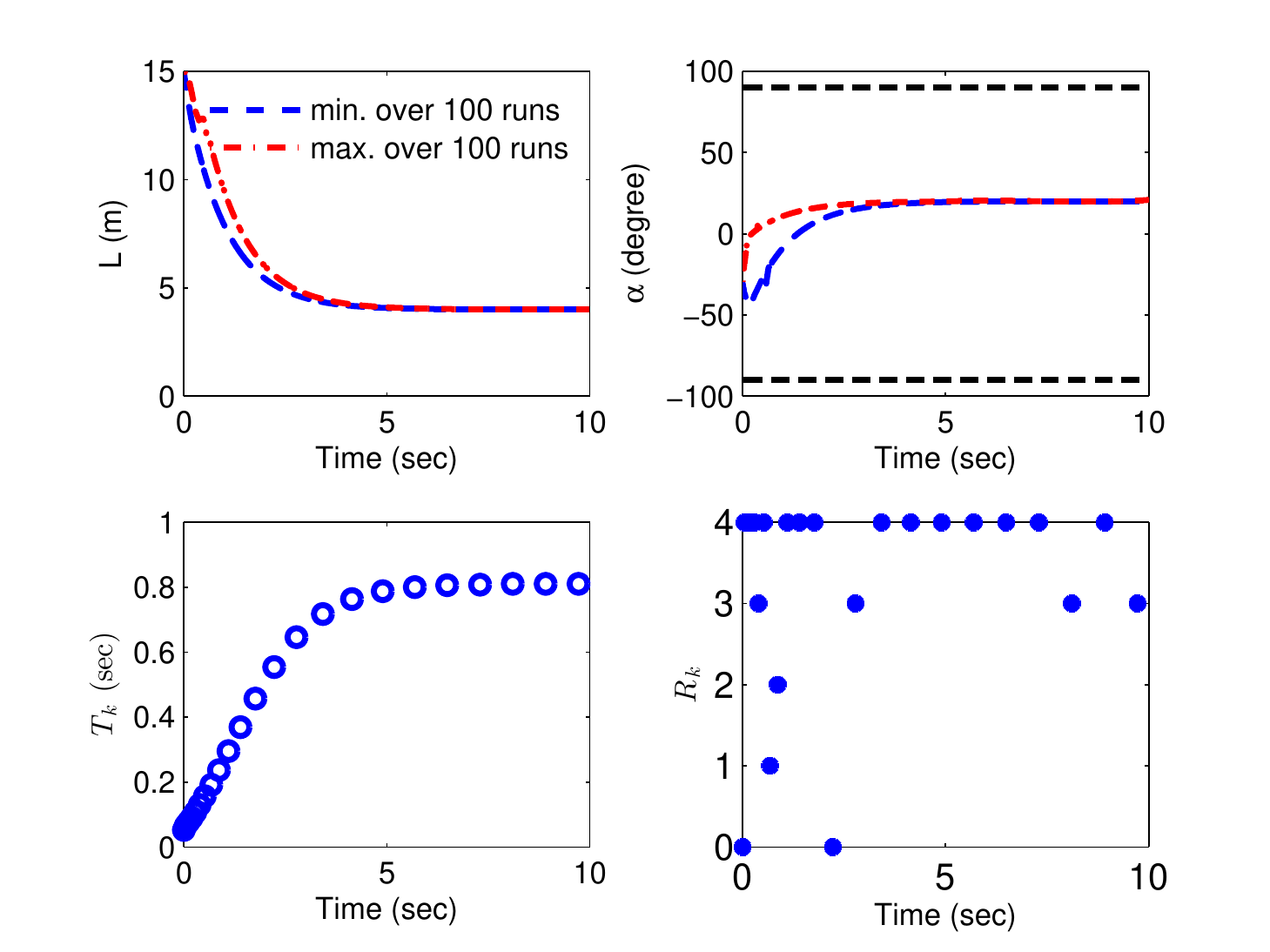}
	\caption{The maximum and minimum trajectories of $L (m)$ and $\alpha (degree)$ under the self-triggered scheme in \eqref{TI} (top plots) and one sample of the inter-transmission time interval $T_k$ and number of received bits $R_{k}$~(bottom plots).}
	\label{fig:states}
\end{figure}

The first simulation is to verify the \emph{almost surely asymptotic stability} of the leader-follower example under the proposed self-triggered scheme in \eqref{TI}~(Theorem \ref{theorem:almost-sure-as}). The upper plots in Fig. \ref{fig:states} show the maximum (red dashed-dot lines) and minimum (blue dashed lines) value of the inter-vehicle distance $L$ and bearing angle $\alpha$ over all the $100$ samples from $0$ to $10$ seconds. From these plots, one can easily see that the maximum and minimum values of the system states  asymptotically converge to the desired set-points $L_d=4~(m)$ and $\alpha_d=20^{\circ}$ as time increases. This is the behavior that one would expect if the system is almost surely asymptotically stable. The lower plots in Fig. \ref{fig:states} shows one sample of the inter-transmission time interval $T_k$~(left plot) and the number of received bits $R_{k}$~(right plot) that are used to achieve the system performance shown in the upper plots.  The transmission time interval $T_k$ is generated by \eqref{TI}. It is clear from the plots that the self-triggered transmission policy starts with small $T_k$ when the leader-follower communication begins in a bad channel region due to a large inter-vehicle distance and bearing angle.  As the leader-follower system gradually approaches its desired formation, the self-triggered communication scheme adaptively increases the inter-transmission time interval to ensure efficient use of communication bandwidth.

The second simulation is to compare the performance of the proposed self-triggered scheme in \eqref{TI} against conventional event-triggered scheme in \cite{wang2011attentively}. For the purpose of comparison, a state dependent event-triggered scheme in \cite{wang2011attentively} was used to trigger the transmission whenever the estimation error exceeded a state dependent threshold. Let $|e(t)|=|\alpha(t)-\hat{\alpha}(t)| \leq 0.1591|[\alpha(t)-\alpha_d, L(t)-L_d]|$ be the triggering condition, and the threshold was selected to assure the same convergent performance as our self-triggered method but in the absence of channel fading. 

\begin{figure}[!ht]
	\centering
	\includegraphics[width=0.53\textwidth]{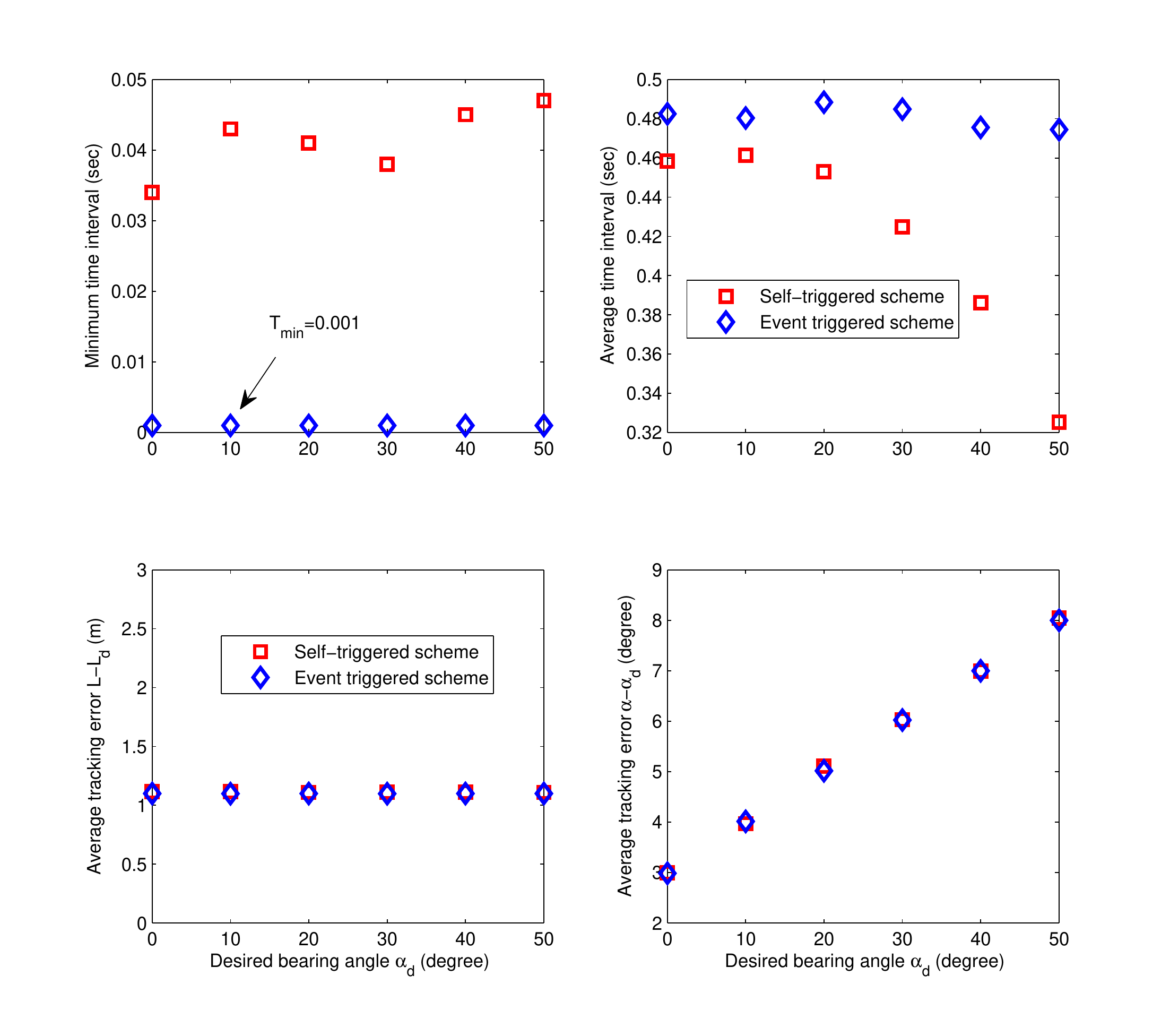}
	\caption{Comparison of minimum transmission time interval~(top left plot), average~(top right plot) transmission time intervals and tracking errors in distance~(bottom left plot) and bearing angle~(bottom right plot) under desired bearing angles $\alpha_d=0^{\circ}, 10^{\circ}, 20^{\circ}, 30^{\circ}, 40^{\circ}, 50^{\circ}$.}
	\label{fig:min-ave-tracking-error}
\end{figure}

Fig. \ref{fig:min-ave-tracking-error} shows the comparison of both transmission time interval and tracking performance for the leader-follower example under the proposed self-triggered scheme~(marked by red squares) in \eqref{TI} and event-triggered scheme~(marked by blue diamonds) in \cite{wang2011attentively} over a wide range of formations, from $\alpha_d=0^{\circ}$ to $\alpha_d=50^{\circ}$. The tracking performance is compared by calculating the expected\footnote{The expectation is approximated by the average of $100$ sample runs, i.e., $\mathbb{E} \frac{1}{10}\int_{0}^{10}|x(t)-x^d|dt \approx \frac{1}{100}\sum_{i=1}^{100}\frac{1}{10}\int_{0}^{10}|x_{i}(t)-x^d|dt$ where $x=[L;\alpha]$ and $x_{i}$ is the $i^{th}$ sample run.} average tracking error of inter-vehicle distance and bearing angles over a time interval $[0, 10]$. The bottom plots of Fig. \ref{fig:min-ave-tracking-error} show that both triggering schemes achieve quite similar tracking performance for inter-vehicle distance $L$ and bearing angle $\alpha$ over all desired formations. The results in the top left plot of Fig. \ref{fig:min-ave-tracking-error} show that the minimum transmission time interval $T_{min}$ that is used to achieve the tracking performance under our proposed self-triggered scheme~(around $T_{min}=0.04$ second) is approximately $40$ times larger than that generated by the event-triggered scheme~($T_{min}=0.001$ second). Note that the minimum transmission time interval determines the channel bandwidth that is actually needed in vehicular networks. This observation implies that our proposed self-triggered scheme allows much more efficient use of communication bandwidth than the traditional event-triggered methods by providing much larger minimum transmission time interval. The comparison of average transmission time intervals under both triggered schemes is provided in the top-right plot of Fig. \ref{fig:min-ave-tracking-error}, which shows that the average interval generated by self-triggered scheme is relatively close to that of the event-triggered one when the desired formations are positioned in good channel regions, such as $\alpha_{d}=0^{\circ}, 10^{\circ}, 20^{\circ}$. When the desired formation configuration approaches bad channel regions, such as $\alpha_d=30^{\circ}, 40^{\circ}, 50^{\circ}$, our proposed self-triggered scheme reacts to those formation changes by adaptively adjusting the average transmission time intervals. As shown in the top right plot of Fig. \ref{fig:min-ave-tracking-error}, the average transmission time interval decreases to ensure sufficient information updates as the desired formations approach bad channel regions.

\begin{figure}[!ht]
	\centering
	\includegraphics[width=0.5\textwidth]{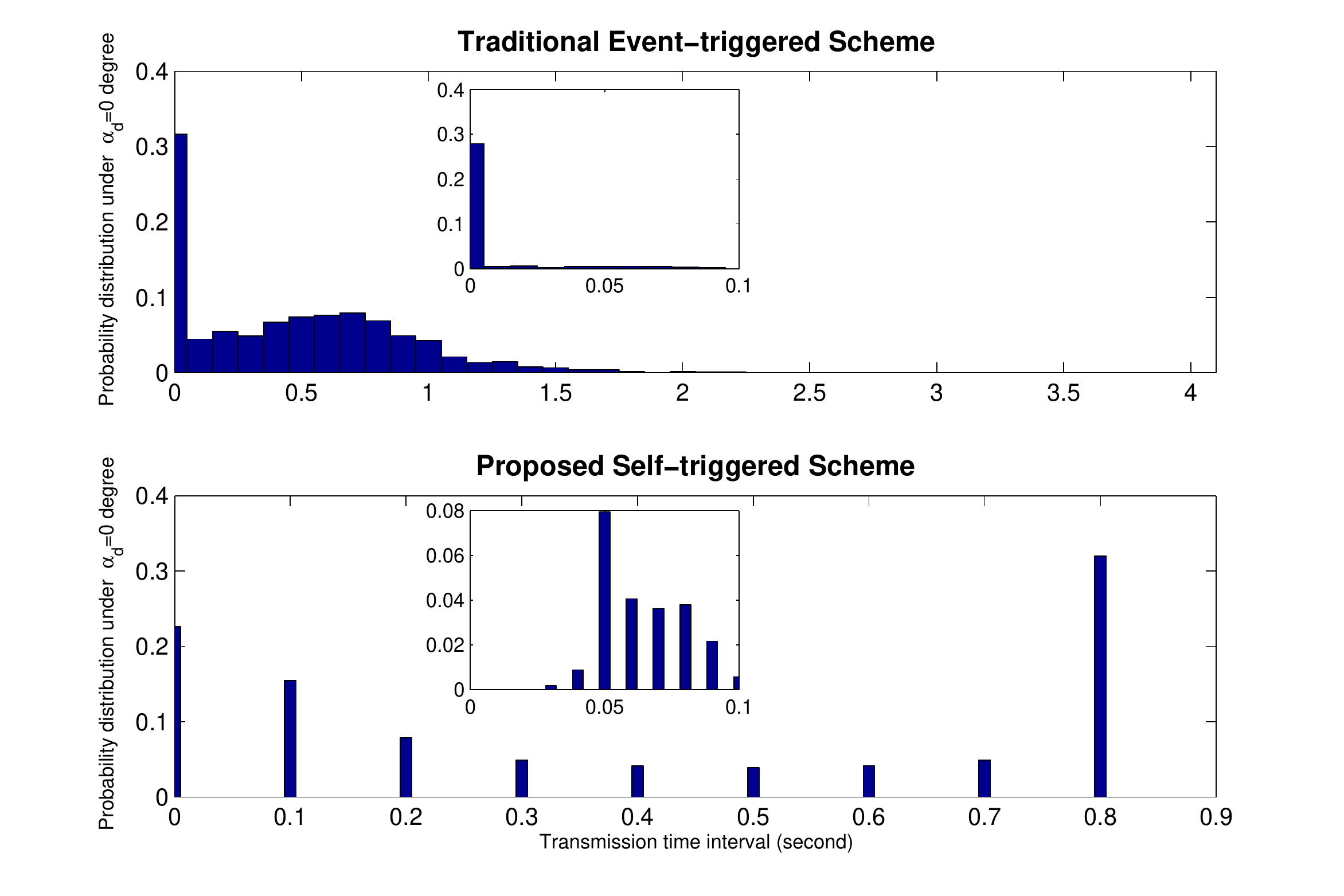}
	\caption{Distribution of inter-transmission time interval under self-triggered scheme in Theorem \ref{theorem:almost-sure-as} and traditional event-triggered scheme \cite{wang2011attentively}.}
	\label{fig:distribution_TT}
\end{figure}

Fig. \ref{fig:distribution_TT} shows the probability distribution of the transmission interval over $100$ runs under the proposed self-triggered scheme (top plot) and traditional event-triggered scheme in \cite{wang2011attentively} (bottom plot) when the desired formation is in good channel region, $\alpha_d=0^{\circ}$. The result shows that even in the good channel region, nearly $30\%$ of the time intervals generated by the event-triggered scheme proposed in \cite{wang1995finite-state}~(top plot in Fig. \ref{fig:distribution_TT}) is below $0.01$ second while the percentage of small time intervals below $0.01$ second in our proposed self-triggered scheme is $0$. This is not surprising since the state-dependent threshold $|e(t)| \leq 0.1591|[\alpha(t)-\alpha_d, L(t)-L_d]|$ in event-triggered scheme, is very sensitive to any small changes on the system states and easy to be violated when they are around the equilibrium. 

\begin{figure}[!ht]
	\centering
	\includegraphics[width=0.5\textwidth]{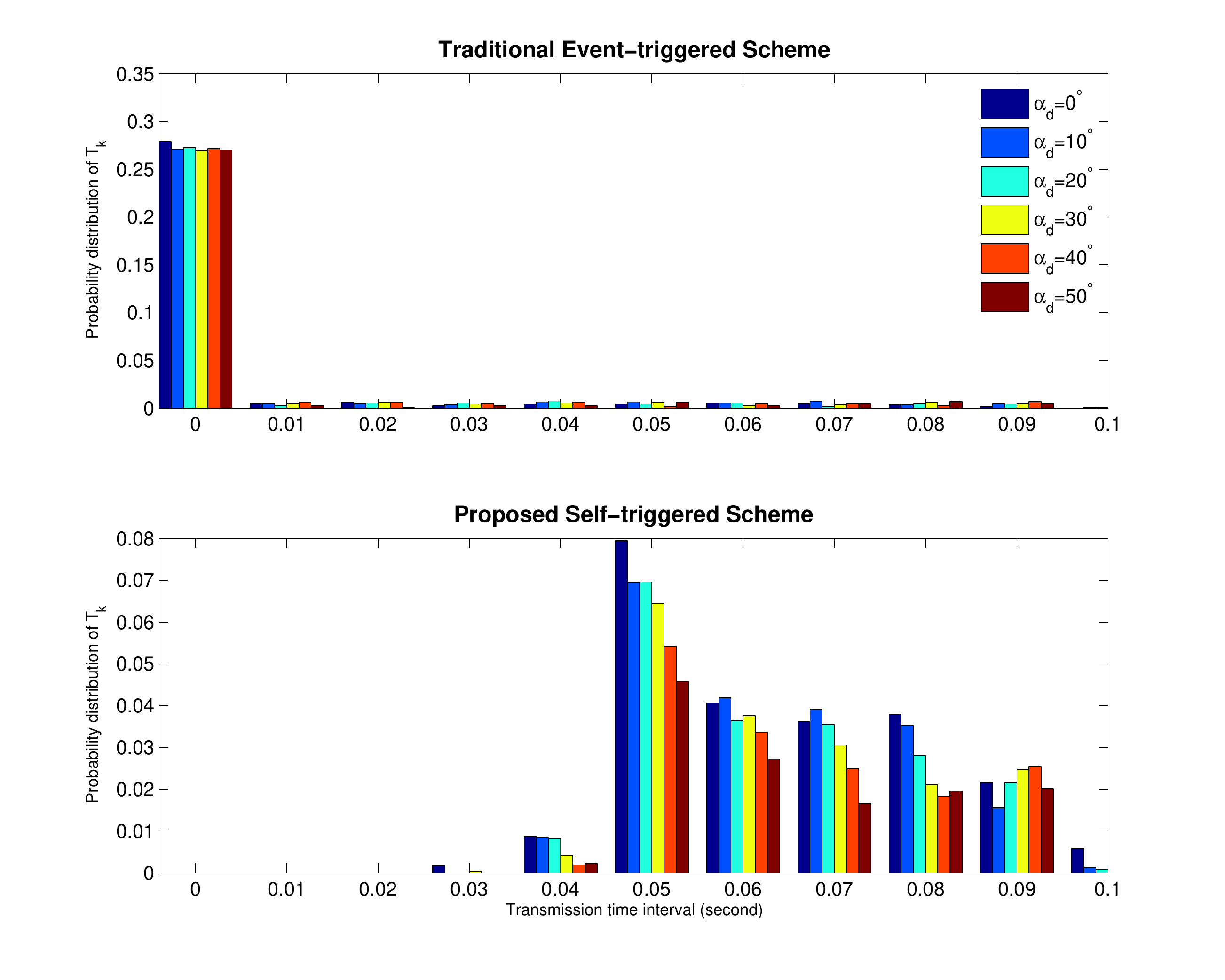}
	\caption{Distribution of small inter-transmission time intervals under self-triggered scheme in \eqref{TI} and event-triggered scheme \cite{wang2011attentively} under desired bearing angles $\alpha_d=0^{\circ}, 10^{\circ}, 20^{\circ}, 30^{\circ}, 40^{\circ}, 50^{\circ}$.}
	\label{fig:distribution-TI-changing-formation}
\end{figure}

In this simulation, we are also interested in testing how robust both triggered schemes are against a wide range of channel fading levels.  The robustness of both triggered schemes is evaluated by examining how frequently a small transmission time interval occurs due to channel fading from $\alpha_d=0^{\circ}$ to $\alpha_d=50^{\circ}$.  Fig. \ref{fig:distribution-TI-changing-formation} shows the probability distributions of the transmission time interval lying in each of the intervals $\cup_{i=0}^{9}[i*0.01, (i+1)*0.01]$ second   under the proposed self-triggered scheme~(bottom plot) and event-triggered scheme~(top plot). The results show that nearly $30\%$ percent of the time intervals generated by the event-triggered scheme in \cite{wang2011attentively} lies in the interval $[0, 0.01]$ second while the percentage generated by the self-triggered scheme in \eqref{TI} is $0$ under all levels of channel fading. This suggests that our proposed self-triggered scheme is more robust against channel variations than the event-triggered scheme. 

\begin{figure}[!ht]
	\centering
	\includegraphics[width=0.53\textwidth]{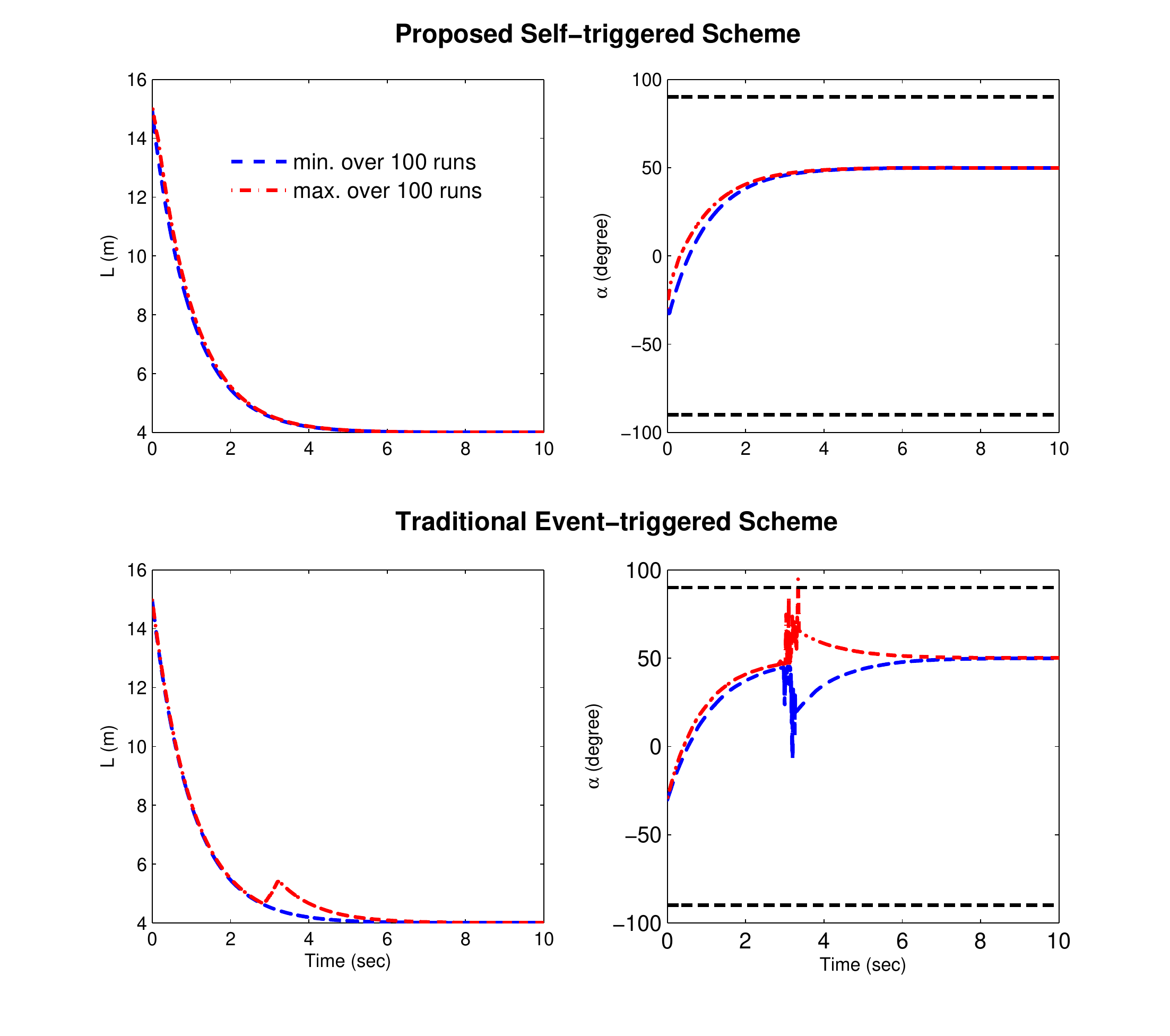}
	\caption{Comparison of state trajectories $L$ and $\alpha$ under self-triggered scheme in Theorem \ref{theorem:almost-sure-as} (top plots) and traditional event-triggered scheme \cite{wang2011attentively} (bottom plots) in the presence of deep fade that lasts for $0.5$ seconds.}
	\label{fig:comp_alpha_L}
\end{figure}

The third simulation considered the deep fading scenario when there was a long consecutive string of dropouts occurring as vehicles approach their desired set-points. The deep fade lasted for $0.6$ seconds. Fig. \ref{fig:comp_alpha_L} shows the comparison of state trajectories $\alpha$ and $L$ over $100$ runs under the self-triggering scheme in Theorem \ref{theorem:almost-sure-as} (top plots) and event-triggered control using threshold $0.1591|[\alpha(t)-\alpha_d, L(t)-L_d]|$ (bottom plots). The plots show that the bearing angle $\alpha$ exhibits a large excursion from the set-point under the traditional event-triggered scheme when a long string of dropouts starts at around $3$ seconds. In contrast, such large deviation of bearing angle $\alpha$ does not occur under the proposed self-triggered scheme, which implies that our proposed self-triggered scheme is more resilient than traditional event-triggered scheme in the presence of severe communication failure.


\subsection{Experimental Results in MobileSim Simulator}
This section compares the system performance and transmission efficiency in the MobileSim Simulator under the traditional event-triggered control \cite{wang2011attentively} and the self-triggered scheme in Theorem \ref{theorem:almost-sure-as}. In the experiments, a sampling time interval is set to $20$ms, which is sufficiently small compared to the dynamics of the robots. Thus, the transmission sequence generated by both the triggered schemes will be evaluated based on the basis of $20$ms. The experiment simulates an interesting and nontrivial scenario when the leader-follower chain is required to travel on a highway with a sharp-curve shape. Such scenario is commonly encountered in the smart transportation system, which is precisely the critical situation that needs wireless communication most to ensure system safety. 

\begin{figure*}[!ht]
\centering
\subfloat[The $1$st second in animation][The $1$st second in animation]
{\includegraphics[width=0.4\textwidth]{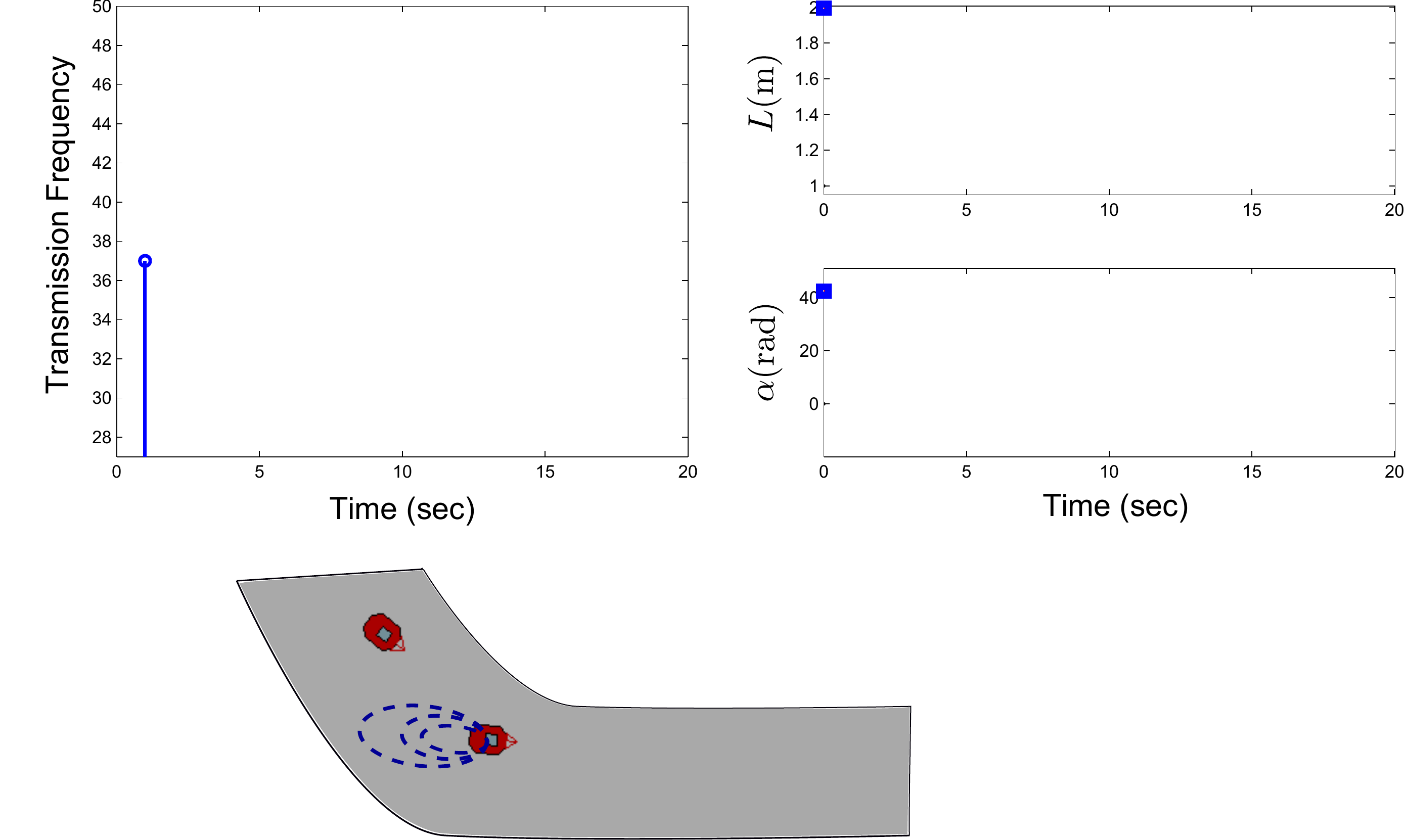}}
\subfloat[The $7$th second in animation][The $7$th second in animation]{
\includegraphics[width=0.4\textwidth]{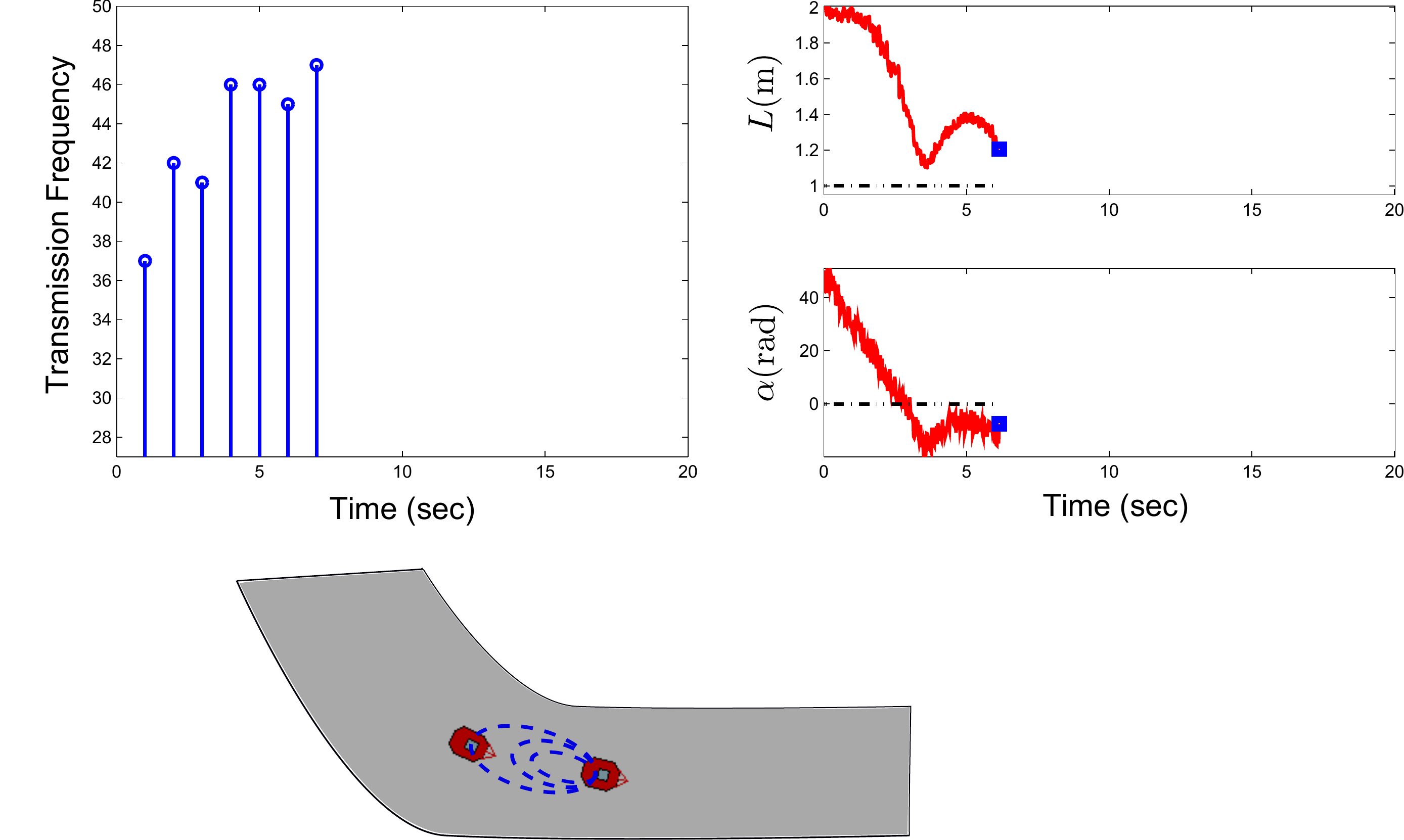}} \\
\subfloat[The $13$th second in animation][The $13$th second in animation]{
	\includegraphics[width=0.4\textwidth]{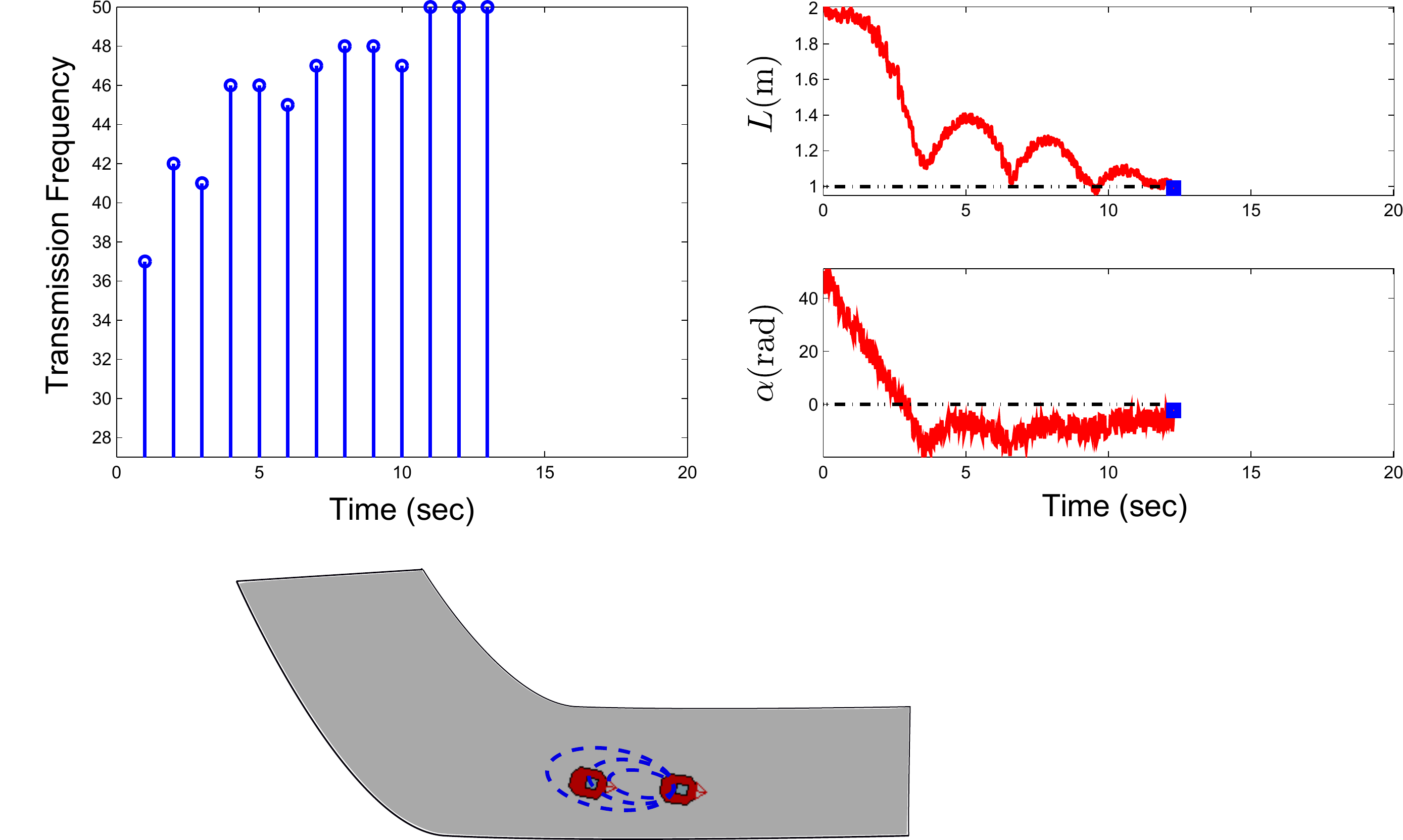}}
\subfloat[The $20$th second in animation][The $20$th second in animation]{
	\includegraphics[width=0.4\textwidth]{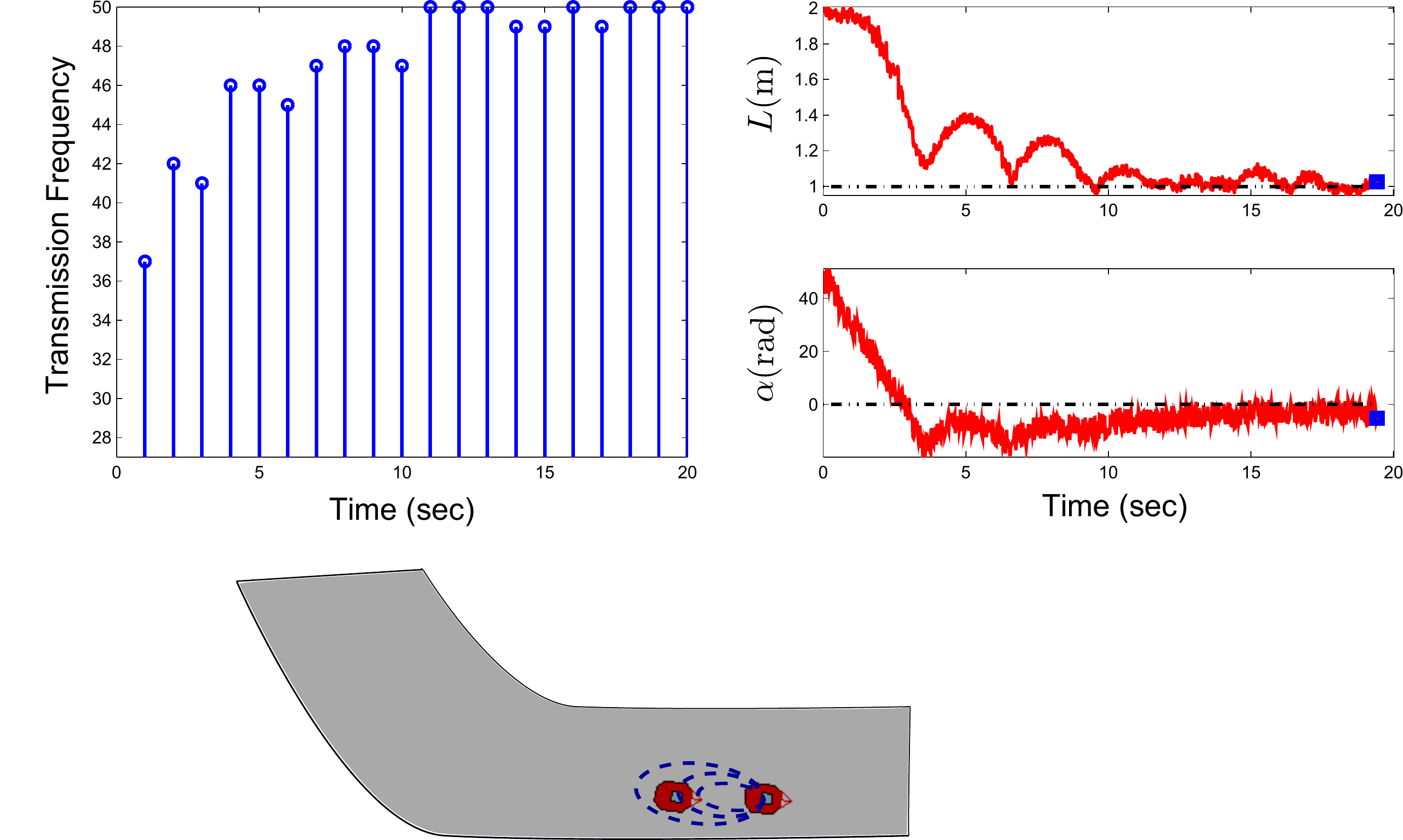}}
\caption{The snapshots of traditional event triggering scheme \cite{wang2011attentively} in the leader-follower formation with $\alpha_{d}=0$ and $L_d=1$ m}
\label{fig: snapshots-event-triggered}
\end{figure*}

\begin{figure*}[!ht]
	\centering
	\subfloat[The $1$st second in animation][The $1$st second in animation]
	{\includegraphics[width=0.4\textwidth]{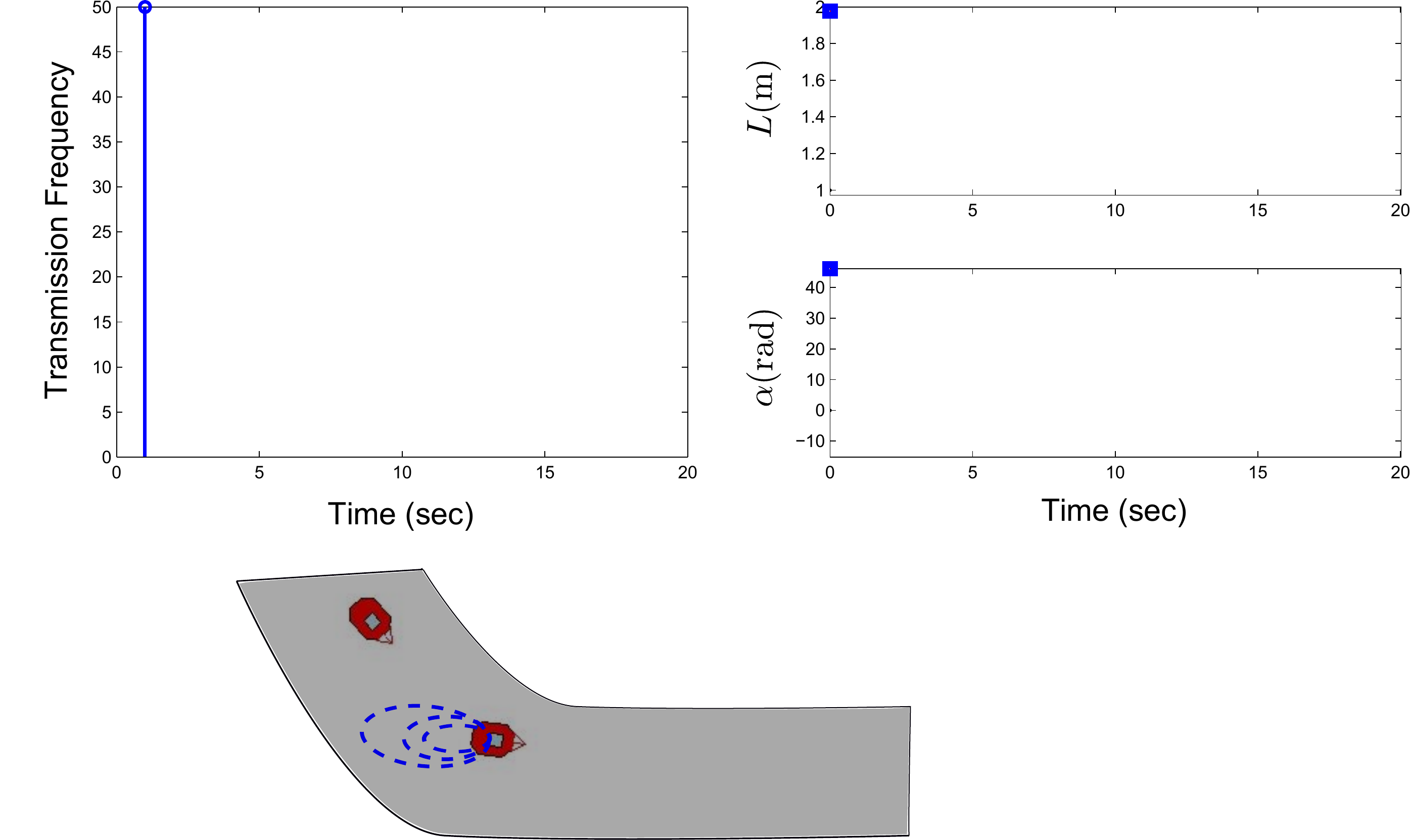}}
	\subfloat[The $7$th second in animation][The $7$th second in animation]{
		\includegraphics[width=0.4\textwidth]{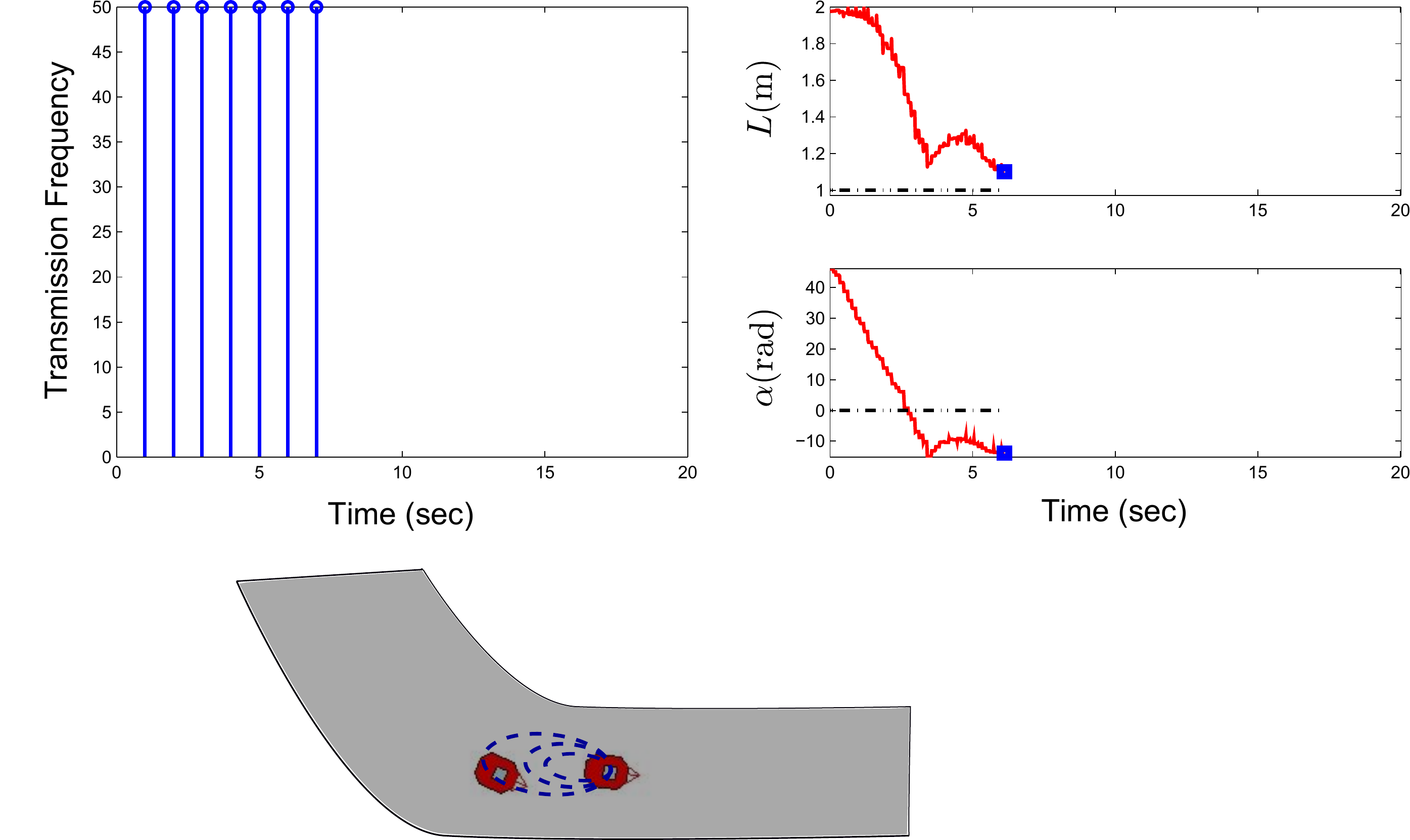}} \\
	\subfloat[The $13$th second in animation][The $13$th second in animation]{
		\includegraphics[width=0.4\textwidth]{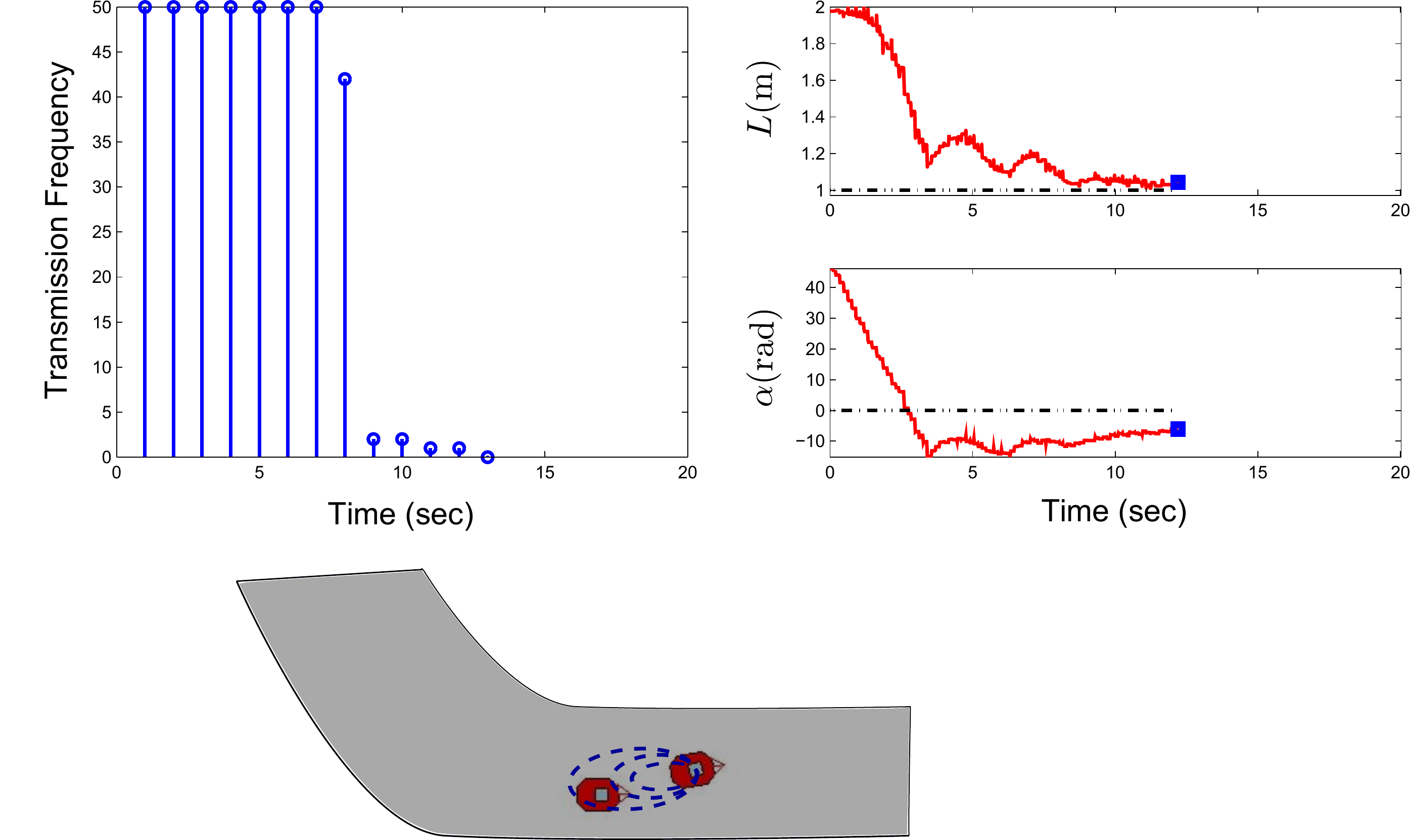}}
	\subfloat[The $20$th second in animation][The $20$th second in animation]{
		\includegraphics[width=0.4\textwidth]{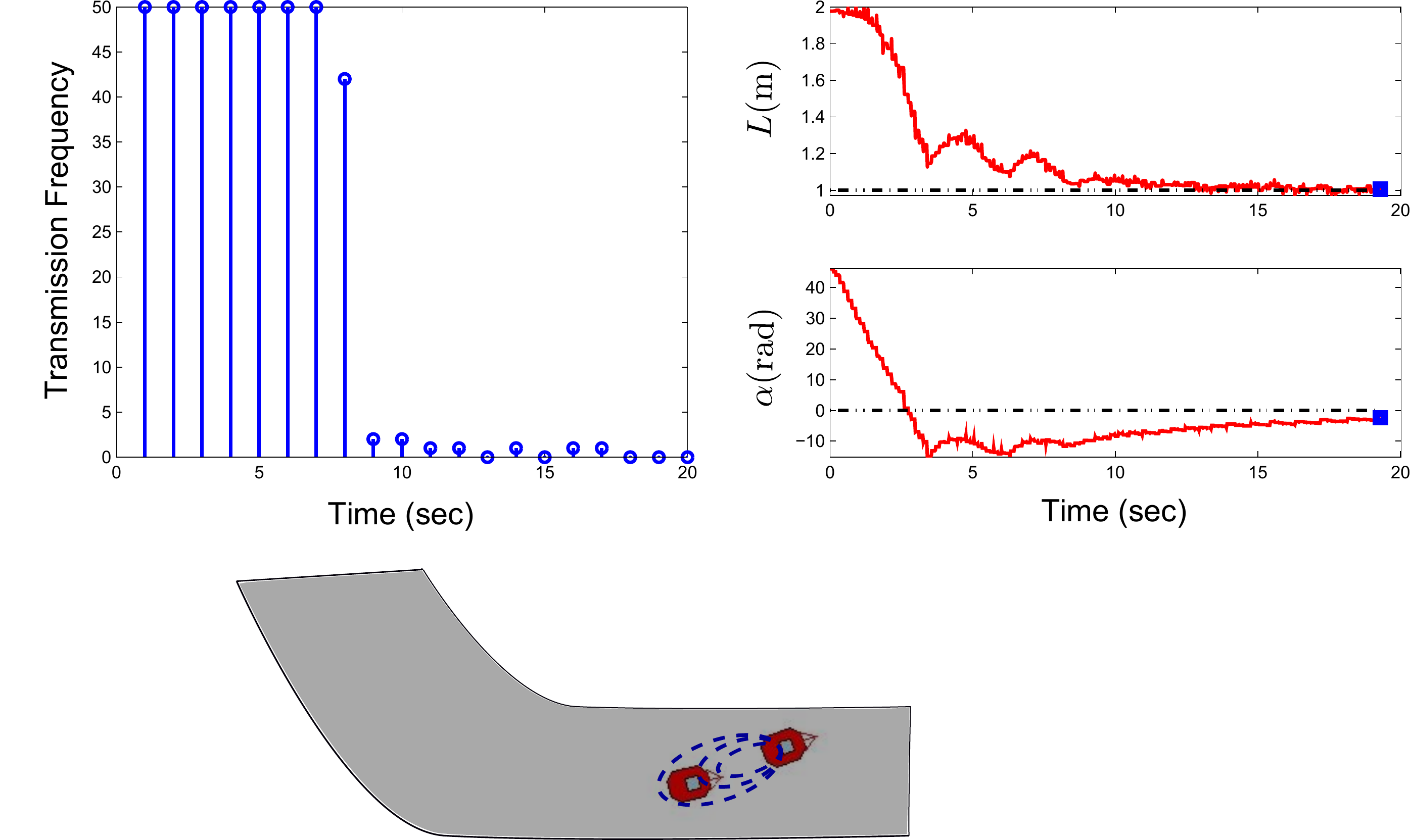}}
	\caption{The snapshots of self-triggering scheme in Theorem \ref{theorem:almost-sure-as} in the leader-follower formation with $\alpha_{d}=0$ and $L_d=1$ m}
	\label{fig: snapshots-self-triggered}
\end{figure*}

The experiments are run from $0$ second to $20$ second. Figs. \ref{fig: snapshots-event-triggered} and \ref{fig: snapshots-self-triggered} are snapshots showing the information, such as state trajectory and transmission frequency, of $1$st, $7$th, $13$th and $20$th second in the experiment animations. In both figures, the upper left plots show the evolutions of the transmission frequency when time progresses while the upper right plots show the corresponding system trajectories regarding the inter-vehicle distance $L$ and relative bearing angle $\alpha$. The bottom plots show the physical locations of the leader-follower chain along the time. As discussed Section \ref{sec: leader-follower}, the leader is equipped with a directional antenna whose radiation pattern is shown in blue dashed ovals in the bottom plots. 

The snapshots in Fig. \ref{fig: snapshots-event-triggered} show that the transmission frequency gradually increases as the leader-follower chain approaches their desired formation under the traditional event-triggered scheme. In particular, the transmission frequency plots show that the channel utilization under event-triggered scheme achieves and remains its maximum limit after $11$ second (transmit every $20$ms). This suggests that the communication channel is constantly occupied while the desired formation is obtained. The inefficient use of communication channels at desired formation will inevitably lead to complete loss of system resilience to any small changes to the system equilibrium. One important cause for the loss of robustness is that the traditional event-triggered control fails to account for the correlation between channel state and physical states in V2V communication. 

In contrast, Fig. \ref{fig: snapshots-self-triggered} shows that the transmission frequency decreases as the desired formation is attained under the self-triggered control proposed in Theorem \ref{theorem:almost-sure-as}. In particular, the transmission frequency plots in Fig. \ref{fig: snapshots-self-triggered} show that the leader-follower chain consumes more channel bandwidth when the vehicles are far apart and less aligned with directional antennae (before the $7$th second) than the time when the vehicles are close-by and aligned with directional antennae (after the $9$th second).  This performance is consistent with the fact that the channel state in V2V is closely related to the physical states of the vehicles. Our proposed self-triggered scheme takes advantages of such relationship to efficiently adjust the communication utilizations. 

\section{Conclusion}
\label{section: conclusion}
This paper developed a novel self-triggered scheme for VNS in the presence of \emph{state-dependent fading channels}. By using a state-dependent fading channel model, the results showed that the proposed self-triggered schemes can achieve efficient use of communication bandwidth with Zeno-free transmission while ensuring four types of \emph{stochastic stability}. Under the proposed self-triggered scheme, this paper also presented a new source coding scheme under which the vehicle's states were tracked with performance guarantee even when the channel states are time varying and stochastically change as a function of vehicle states. Both simulation and experimental results demonstrated that the proposed self-triggered scheme was more efficient in bandwidth utilization and more resilient to deep fades than traditional event-triggered schemes. 

\bibliographystyle{IEEEtran}        
\bibliography{bibfile/TAC-2017} 

\begin{thebibliography}{10}
\providecommand{\url}[1]{#1}
\csname url@samestyle\endcsname
\providecommand{\newblock}{\relax}
\providecommand{\bibinfo}[2]{#2}
\providecommand{\BIBentrySTDinterwordspacing}{\spaceskip=0pt\relax}
\providecommand{\BIBentryALTinterwordstretchfactor}{4}
\providecommand{\BIBentryALTinterwordspacing}{\spaceskip=\fontdimen2\font plus
\BIBentryALTinterwordstretchfactor\fontdimen3\font minus
  \fontdimen4\font\relax}
\providecommand{\BIBforeignlanguage}[2]{{%
\expandafter\ifx\csname l@#1\endcsname\relax
\typeout{** WARNING: IEEEtran.bst: No hyphenation pattern has been}%
\typeout{** loaded for the language `#1'. Using the pattern for}%
\typeout{** the default language instead.}%
\else
\language=\csname l@#1\endcsname
\fi
#2}}
\providecommand{\BIBdecl}{\relax}
\BIBdecl

\bibitem{papadimitratos2009vehicular}
P.~Papadimitratos, A.~De~La~Fortelle, K.~Evenssen, R.~Brignolo, and S.~Cosenza,
  ``Vehicular communication systems: Enabling technologies, applications, and
  future outlook on intelligent transportation,'' \emph{IEEE Communications
  Magazine}, vol.~47, no.~11, pp. 84--95, 2009.

\bibitem{kenney2011dedicated}
J.~B. Kenney, ``Dedicated short-range communications (dsrc) standards in the
  united states,'' \emph{Proceedings of the IEEE}, vol.~99, no.~7, pp.
  1162--1182, 2011.

\bibitem{cheng2007mobile}
L.~Cheng, B.~E. Henty, D.~D. Stancil, F.~Bai, and P.~Mudalige, ``Mobile
  vehicle-to-vehicle narrow-band channel measurement and characterization of
  the 5.9 ghz dedicated short range communication (dsrc) frequency band,''
  \emph{IEEE Journal on Selected Areas in Communications}, vol.~25, no.~8, pp.
  1501--1516, 2007.

\bibitem{park2014high}
P.~Park, H.~Khadilkar, H.~Balakrishnan, and C.~J. Tomlin, ``High confidence
  networked control for next generation air transportation systems,''
  \emph{Automatic Control, IEEE Transactions on}, vol.~59, no.~12, pp.
  3357--3372, 2014.

\bibitem{park2012investigating}
P.~Park and C.~Tomlin, ``Investigating communication infrastructure of next
  generation air traffic management,'' in \emph{Proceedings of the 2012
  IEEE/ACM Third International Conference on Cyber-Physical Systems}.\hskip 1em
  plus 0.5em minus 0.4em\relax IEEE Computer Society, 2012, pp. 35--44.

\bibitem{akyildiz2005underwater}
I.~F. Akyildiz, D.~Pompili, and T.~Melodia, ``Underwater acoustic sensor
  networks: research challenges,'' \emph{Ad hoc networks}, vol.~3, no.~3, pp.
  257--279, 2005.

\bibitem{partan2007survey}
J.~Partan, J.~Kurose, and B.~N. Levine, ``A survey of practical issues in
  underwater networks,'' \emph{ACM SIGMOBILE Mobile Computing and
  Communications Review}, vol.~11, no.~4, pp. 23--33, 2007.

\bibitem{tse2005fundamentals}
D.~Tse and P.~Viswanath, \emph{Fundamentals of wireless communication}.\hskip
  1em plus 0.5em minus 0.4em\relax Cambridge university press, 2005.

\bibitem{wang2011event}
X.~Wang and M.~D. Lemmon, ``Event-triggering in distributed networked control
  systems,'' \emph{Automatic Control, IEEE Transactions on}, vol.~56, no.~3,
  pp. 586--601, 2011.

\bibitem{guinaldo2012distributed}
M.~Guinaldo, D.~Lehmann, J.~Sanchez, S.~Dormido, and K.~H. Johansson,
  ``Distributed event-triggered control with network delays and packet
  losses,'' in \emph{2012 IEEE 51st Annual Conference on Decision and Control
  (CDC)}.\hskip 1em plus 0.5em minus 0.4em\relax IEEE, 2012, pp. 1--6.

\bibitem{hu2014event}
B.~Hu and M.~D. Lemmon, ``Event triggering in vehicular networked systems with
  limited bandwidth and deep fading,'' in \emph{Decision and Control (CDC),
  2014 IEEE 53rd Annual Conference on}.\hskip 1em plus 0.5em minus 0.4em\relax
  IEEE, 2014, pp. 3542--3547.

\bibitem{ploeg2014lp}
J.~Ploeg, N.~Van De~Wouw, and H.~Nijmeijer, ``$\mathcal{L}_p$ string stability
  of cascaded systems: Application to vehicle platooning,'' \emph{IEEE
  Transactions on Control Systems Technology}, vol.~22, no.~2, pp. 786--793,
  2014.

\bibitem{tanner2004leader}
H.~G. Tanner, G.~J. Pappas, and V.~Kumar, ``Leader-to-formation stability,''
  \emph{IEEE Transactions on robotics and automation}, vol.~20, no.~3, pp.
  443--455, 2004.

\bibitem{tomlin2000game}
C.~J. Tomlin, J.~Lygeros, and S.~S. Sastry, ``A game theoretic approach to
  controller design for hybrid systems,'' \emph{Proceedings of the IEEE},
  vol.~88, no.~7, pp. 949--970, 2000.

\bibitem{wang2009self}
X.~Wang and M.~D. Lemmon, ``Self-triggered feedback control systems with
  finite-gain $\mathcal{L}_{2}$ stability,'' \emph{IEEE transactions on
  automatic control}, vol.~54, no.~3, pp. 452--467, 2009.

\bibitem{tabuada2007event}
P.~Tabuada, ``Event-triggered real-time scheduling of stabilizing control
  tasks,'' \emph{IEEE Transactions on Automatic Control}, vol.~52, no.~9, pp.
  1680--1685, 2007.

\bibitem{hetel2017recent}
L.~Hetel, C.~Fiter, H.~Omran, A.~Seuret, E.~Fridman, J.-P. Richard, and S.~I.
  Niculescu, ``Recent developments on the stability of systems with aperiodic
  sampling: An overview,'' \emph{Automatica}, vol.~76, pp. 309--335, 2017.

\bibitem{heemels2012introduction}
W.~Heemels, K.~H. Johansson, and P.~Tabuada, ``An introduction to
  event-triggered and self-triggered control,'' in \emph{Decision and Control
  (CDC), 2012 IEEE 51st Annual Conference on}.\hskip 1em plus 0.5em minus
  0.4em\relax IEEE, 2012, pp. 3270--3285.

\bibitem{walsh2002stability}
G.~C. Walsh, H.~Ye, and L.~G. Bushnell, ``Stability analysis of networked
  control systems,'' \emph{IEEE transactions on control systems technology},
  vol.~10, no.~3, pp. 438--446, 2002.

\bibitem{zhang2001stability}
W.~Zhang, M.~S. Branicky, and S.~M. Phillips, ``Stability of networked control
  systems,'' \emph{IEEE Control Systems}, vol.~21, no.~1, pp. 84--99, 2001.

\bibitem{lehmann2012event}
D.~Lehmann and J.~Lunze, ``Event-based control with communication delays and
  packet losses,'' \emph{International Journal of Control}, vol.~85, no.~5, pp.
  563--577, 2012.

\bibitem{wu2014formal}
B.~Wu, H.~Lin, and M.~Lemmon, ``Formal methods for stability analysis of
  networked control systems with {IEEE} 802.15. 4 protocol,'' in \emph{Decision
  and Control (CDC), 2014 IEEE 53rd Annual Conference on}.\hskip 1em plus 0.5em
  minus 0.4em\relax IEEE, 2014, pp. 5266--5271.

\bibitem{dolk2017event}
V.~Dolk and M.~Heemels, ``Event-triggered control systems under packet
  losses,'' \emph{Automatica}, vol.~80, pp. 143--155, 2017.

\bibitem{yu2013event}
H.~Yu and P.~J. Antsaklis, ``Event-triggered output feedback control for
  networked control systems using passivity: Achieving l2 stability in the
  presence of communication delays and signal quantization,''
  \emph{Automatica}, vol.~49, no.~1, pp. 30--38, 2013.

\bibitem{peng2013event}
C.~Peng and T.~C. Yang, ``Event-triggered communication and h∞ control
  co-design for networked control systems,'' \emph{Automatica}, vol.~49, no.~5,
  pp. 1326--1332, 2013.

\bibitem{dolk2017output}
V.~Dolk, D.~P. Borgers, and W.~Heemels, ``Output-based and decentralized
  dynamic event-triggered control with guaranteed $\mathcal{L}_p$-gain
  performance and zeno-freeness,'' \emph{IEEE Transactions on Automatic
  Control}, vol.~62, no.~1, pp. 34--49, 2017.

\bibitem{akki1994statistical}
A.~S. Akki, ``Statistical properties of mobile-to-mobile land communication
  channels,'' \emph{IEEE transactions on vehicular technology}, vol.~43, no.~4,
  pp. 826--831, 1994.

\bibitem{akki1986statistical}
A.~S. Akki and F.~Haber, ``A statistical model of mobile-to-mobile land
  communication channel,'' \emph{IEEE transactions on vehicular technology},
  vol.~35, no.~1, pp. 2--7, 1986.

\bibitem{borgers2014event}
D.~N. Borgers and W.~M. Heemels, ``Event-separation properties of
  event-triggered control systems,'' \emph{IEEE Transactions on Automatic
  Control}, vol.~59, no.~10, pp. 2644--2656, 2014.

\bibitem{li2016event}
H.~Li, Z.~Chen, L.~Wu, and H.-K. Lam, ``Event-triggered control for nonlinear
  systems under unreliable communication links,'' \emph{IEEE Transactions on
  Fuzzy Systems}, 2016.

\bibitem{mazo2008event}
M.~Mazo and P.~Tabuada, ``On event-triggered and self-triggered control over
  sensor/actuator networks,'' in \emph{Decision and Control, 2008. CDC 2008.
  47th IEEE Conference on}.\hskip 1em plus 0.5em minus 0.4em\relax IEEE, 2008,
  pp. 435--440.

\bibitem{heemels2013periodic}
W.~Heemels, M.~Donkers, and A.~R. Teel, ``Periodic event-triggered control for
  linear systems,'' \emph{Automatic Control, IEEE Transactions on}, vol.~58,
  no.~4, pp. 847--861, 2013.

\bibitem{antunes2014rollout}
D.~Antunes and W.~Heemels, ``Rollout event-triggered control: Beyond periodic
  control performance,'' \emph{IEEE Transactions on Automatic Control},
  vol.~59, no.~12, pp. 3296--3311, 2014.

\bibitem{anderson2015self}
R.~P. Anderson, D.~Milutinovi{\'c}, and D.~V. Dimarogonas, ``Self-triggered
  sampling for second-moment stability of state-feedback controlled sde
  systems,'' \emph{Automatica}, vol.~54, pp. 8--15, 2015.

\bibitem{gommans2014self}
T.~Gommans, D.~Antunes, T.~Donkers, P.~Tabuada, and M.~Heemels,
  ``Self-triggered linear quadratic control,'' \emph{Automatica}, vol.~50,
  no.~4, pp. 1279--1287, 2014.

\bibitem{bin2013}
B.~Hu and M.~D. Lemmon, ``Using channel state feedback to achieve resilience to
  deep fades in wireless networked control systems,'' in \emph{Proceedings of
  the 2nd international conference on High Confidence Networked Systems}, April
  9-11 2013.

\bibitem{2015Hu}
B.~Hu and M.~Lemmon, ``Distributed switching control to achieve almost sure
  safety for leader-follower vehicular networked systems,'' \emph{Automatic
  Control, IEEE Transactions on}, vol.~PP, no.~99, pp. 1--1, 2015.

\bibitem{li2017efficiently}
L.~Li, X.~Wang, and M.~D. Lemmon, ``Efficiently attentive event-triggered
  systems with limited bandwidth,'' \emph{IEEE Transactions on Automatic
  Control}, vol.~62, no.~3, pp. 1491--1497, 2017.

\bibitem{Mobilesim}
\BIBentryALTinterwordspacing
\emph{The Adept MobileRobots Simulator}. [Online]. Available:
  \url{http://robots.mobilerobots.com/MobileSim/download/current/}
\BIBentrySTDinterwordspacing

\bibitem{hu2015distributed}
B.~Hu and M.~D. Lemmon, ``Distributed switching control to achieve almost sure
  safety for leader-follower vehicular networked systems,'' \emph{IEEE
  Transactions on Automatic Control}, vol.~60, no.~12, pp. 3195--3209, 2015.

\bibitem{martins2006feedback}
N.~C. Martins, M.~A. Dahleh, and N.~Elia, ``Feedback stabilization of uncertain
  systems in the presence of a direct link,'' \emph{IEEE Transactions on
  Automatic Control}, vol.~51, no.~3, pp. 438--447, 2006.

\bibitem{choudhury2002using}
R.~R. Choudhury, X.~Yang, R.~Ramanathan, and N.~H. Vaidya, ``Using directional
  antennas for medium access control in ad hoc networks,'' in \emph{Proceedings
  of the 8th annual international conference on Mobile computing and
  networking}.\hskip 1em plus 0.5em minus 0.4em\relax ACM, 2002, pp. 59--70.

\bibitem{yi2003capacity}
S.~Yi, Y.~Pei, and S.~Kalyanaraman, ``On the capacity improvement of ad hoc
  wireless networks using directional antennas,'' in \emph{Proceedings of the
  4th ACM international symposium on Mobile ad hoc networking \&
  computing}.\hskip 1em plus 0.5em minus 0.4em\relax ACM, 2003, pp. 108--116.

\bibitem{balanis2016antenna}
C.~A. Balanis, \emph{Antenna theory: analysis and design}.\hskip 1em plus 0.5em
  minus 0.4em\relax John Wiley \& Sons, 2016.

\bibitem{stuber2011principles}
G.~L. St{\"u}ber, \emph{Principles of mobile communication}.\hskip 1em plus
  0.5em minus 0.4em\relax Springer Science \& Business Media, 2011.

\bibitem{nesic2009unified}
D.~Nesic and D.~Liberzon, ``A unified framework for design and analysis of
  networked and quantized control systems,'' \emph{IEEE Transactions on
  Automatic control}, vol.~54, no.~4, pp. 732--747, 2009.

\bibitem{liberzon2005stabilization}
D.~Liberzon and J.~P. Hespanha, ``Stabilization of nonlinear systems with
  limited information feedback,'' \emph{IEEE Transactions on Automatic
  Control}, vol.~50, no.~6, pp. 910--915, 2005.

\bibitem{kozin1969survey}
F.~Kozin, ``A survey of stability of stochastic systems,'' \emph{Automatica},
  vol.~5, no.~1, pp. 95--112, 1969.

\bibitem{Kushner1967}
H.~Kushner, \emph{Stochastic Stability and Control}.\hskip 1em plus 0.5em minus
  0.4em\relax Academic Press, 1967.

\bibitem{nevsic2004input}
D.~Ne{\v{s}}i{\'c} and A.~R. Teel, ``Input-output stability properties of
  networked control systems,'' \emph{Automatic Control, IEEE Transactions on},
  vol.~49, no.~10, pp. 1650--1667, 2004.

\bibitem{bin2014-2}
B.~Hu and M.~D. Lemmon, ``Distributed switching control to achieve resilience
  to deep fades in leader-follower nonholonomic systems,'' in \emph{Proceedings
  of the 3rd international conference on High Confidence Networked Systems},
  April 15-17 2014.

\bibitem{wang2011attentively}
X.~Wang and M.~Lemmon, ``Attentively efficient controllers for event-triggered
  feedback systems,'' in \emph{Decision and Control and European Control
  Conference (CDC-ECC), 2011 50th IEEE Conference on}.\hskip 1em plus 0.5em
  minus 0.4em\relax IEEE, 2011, pp. 4698--4703.

\bibitem{zhang1999finite}
Q.~Zhang and S.~A. Kassam, ``Finite-state markov model for rayleigh fading
  channels,'' \emph{IEEE Transactions on communications}, vol.~47, no.~11, pp.
  1688--1692, 1999.

\bibitem{wang1995finite-state}
H.~Wang and N.~Moayeri, ``Finite-state markov channel - a useful model for
  radio communication channels,'' \emph{IEEE Transactions on Vehicular
  Technology}, vol.~44, no.~1, pp. 163--171, 1995.

\bibitem{jiang2001input}
Z.-P. Jiang and Y.~Wang, ``Input-to-state stability for discrete-time nonlinear
  systems,'' \emph{Automatica}, vol.~37, no.~6, pp. 857--869, 2001.

\end{thebibliography}

\appendix
\label{appendix}
\begin{IEEEproof}[Proof of Theorem \ref{theorem:as-in-expectation}]
The proof is based on the small-gain theorem \cite{jiang2001input}. Let $t_{k}$ denote the time instant for the $k^{th}$ transmission event and consider the dynamic evolution of the estimation over the time interval $[t_{k}, t_{k+1})$. Since Assumption \ref{assumption: e} holds, one has
\begin{align}
W(e) \leq e^{L_{e}(t-t_{k})}W(e(t_{k}^{+}))+L_{x}(e^{L_{e}(t-t_{k})}-1)|\overline{x}|_{[t_{k}, t)}
\end{align}
Since $w_{1}|e| \leq W(e) \leq w_{2}|e|$, one further has
\begin{align*}
|e|_{[t_{k}, t)}=|e(t)| \leq \frac{w_{2}}{w_{1}}e^{L_{e}(t-t_{k})} |e(t_{k}^{+})| + \frac{L_{x}}{w_{1}L_{e}}(e^{L_{e}(t-t_{k})}-1)|\overline{x}|_{[t_{k}, t)}
\end{align*}
Taking the expectation on both sides of the above inequality yields
\begin{align*}
&\mathbb{E}(|e(t)|) \\
&\leq \frac{w_{2}}{w_{1}}e^{L_{e}(t-t_{k})}\mathbb{E}(2^{-R(k)})\mathbb{E}(|e(t_{k})|) +\frac{L_{x}(e^{L_{e}(t-t_{k})}-1)}{w_{1}L_{e}}\mathbb{E}(|\overline{x}|_{[t_{k}, t)}) \\
&\leq \frac{w_{2}}{w_{1}}e^{L_{e}(t-t_{k})} G_k \mathbb{E}(|e(t_{k})|) +\frac{L_{x}(e^{L_{e}(t-t_{k})}-1)}{w_{1}L_{e}}\mathbb{E}(|\overline{x}|_{[t_{k}, t)})
\end{align*}
where $G_k := G(|x(t_k)|)$. The first inequality holds due to the quantization, $|e(t_{k}^{+})|=2^{-R_k}|e(t_k)|$ and the fact that the random variable $R(k)$ at time $t_k$ is independent of $e(t_k)$ (before the jump). The second inequality holds because of the technical Lemma \ref{lem:G}. Suppose the next transmission time instant $t_{k+1}$ is generated by equation (\ref{TI}), then one has
\begin{align*}
&\mathbb{E}(|e(t_{k+1})|) \\
&\leq \underbrace{\frac{G_{k}\frac{w_{2}}{w_{1}}(1+\frac{L_{x}\overline{\chi}_{1}}{L_{e}w_{1}})}{G_{k}\frac{w_{2}}{w_{1}}+\frac{L_{x}\overline{\chi}_{1}}{L_{e}w_{1}}}}_{< 1}\mathbb{E}(|e(t_{k})|)+\frac{L_{x}}{w_{1}L_{e}}\frac{1-G_{k}\frac{w_{2}}{w_{1}}}{G_{k}\frac{w_{2}}{w_{1}}+\frac{L_{x}\overline{\chi}_{1}}{L_{e}w_{1}}}\mathbb{E}(|\overline{\overline{x}}|_{[t_{k}, t_{k+1})})
\end{align*}
for all $x \in \Omega_{x}=\{ x \in \mathbb{R}^{n} \vert  G(|x|) < w_{1}/w_{2}\}$.  Similarly, by Assumption \ref{assumption: x}, one has
\begin{align*}
\mathbb{E}(|\overline{x}(t)|) &\leq \mathbb{E}(\beta(|\overline{x}(t_k)|, t-t_k)) + \overline{\chi}_{1}\mathbb{E}(|e|_{[t_k, t)}) \\
& \leq \beta(\mathbb{E}(|\overline{x}(t_k)|), t-t_k)+ \overline{\chi}_{1}\mathbb{E}(|e|_{[t_k, t)})
\end{align*}
where the class $\mathcal{KL}$ function $\beta(s, t)$ is concave with respect to $s$, and thus the second inequality holds due to the Jensen's inequality. Since
\begin{equation}
\label{ineq: small-gain}
\overline{\chi}_{1}\cdot \frac{L_{x}}{w_{1}L_{e}}\frac{1-G_{k}\frac{w_{2}}{w_{1}}}{G_{k}\frac{w_{2}}{w_{1}}+\frac{L_{x}\overline{\chi}_{1}}{L_{e}w_{1}}} =\frac{L_x \overline{\chi}_{1}}{L_{e}G_{k}w_{2}+L_x \overline{\chi}_{1}}(1-G_k \frac{w_2}{w_1})< 1
\end{equation}
and $k \in \mathbb{Z}_{\geq 0}$ is arbitrarily chosen, one knows the system with states $\mathbb{E}(|\overline{x}|)$ and $\mathbb{E}(|e(t)|)$ is asymptotically stable by the small-gain theorem, i.e.
\begin{align*}
\lim_{t \rightarrow \infty}\mathbb{E}(|\overline{x}(t)|) \rightarrow 0
\end{align*}
The stability argument is therefore proved.  

The Zeno-free transmission generated by equation \eqref{TI} can also be proved by considering that $\forall x \in \Omega_{x}$
\begin{align}
\label{inside-TI}
\frac{1-\frac{w_{2}}{w_{1}}G( |x|)}{\frac{w_{2}}{w_{1}}G(|x|)+\frac{L_{x}\overline{\chi}_{1}}{L_{e}w_{1}}} > 0
\end{align}
holds. This leads to a strictly positive transmission time interval defined by equation \eqref{TI}. Since the function $G(|x|)$ monotonically increases w.r.t. the state $|x|$, then one knows that the function in \eqref{inside-TI} monotonically decreases w.r.t. $|x|$. Thus, the transmission time interval $T$ generated by \eqref{TI} monotonically decreases w.r.t. $|x|$.   The proof is complete. 
\end{IEEEproof}
\begin{IEEEproof}[Proof of Theorem \ref{theorem:almost-sure-as}]
Following the proof techniques used in Theorem \ref{theorem:as-in-expectation}, one can obtain that the VNS in \eqref{shs} is exponentially stable in expectation with respect to origin under the Exp-ISS assumption stated in Assumption \ref{assumption: x}. Specifically, there must exist an exp-$\mathcal{KL}$ function $\beta(s, t)=c_{1}\exp(-c_2t)s$ such that $\forall \overline{x}(0)$, $\mathbb{E}(|\overline{x}(t)|) \leq c_{1}\exp(-c_{2} t) |\overline{x}(0)|, \forall t \in \mathbb{R}_{\geq 0}$. To prove the almost surely asymptotic stability, let $\tau'> \tau \geq 0$ denote any time instant such that  $\tau \leq t \leq \tau'$ holds, for any given $\epsilon' > 0$, consider the following probability bound,
\begin{align}
{\rm Pr}\{&\sup_{\tau \leq t < \tau'} |\overline{x}(t)| \geq \epsilon'\} \leq \mathbb{E}\big\{\sup_{\tau \leq t < \tau'} |\overline{x}(t)| \big\}/\epsilon' \nonumber \\
 &\leq \mathbb{E}\big\{\int_{\tau}^{\tau'} |\overline{x}(t)|dt \big\}/\epsilon'
 \leq \int_{\tau}^{\tau'} \mathbb{E}\big\{ |\overline{x}(t)| \big\}dt/\epsilon' \nonumber \\
 & \leq \int_{\tau}^{\tau'} c_{1}\exp(-c_{2} t) |\overline{x}(0)|dt /\epsilon' 
 \leq \frac{c_1 |\overline{x}(0)|}{c_2 \epsilon'} \big[e^{-c_{2}\tau}-e^{-c_2 \tau'} \big]
 \label{ineq:probability-bound}
\end{align}
where the first inequality holds due to the Markov inequality and the third inequality holds by exchanging the expectation and integration due to the measurability of the solution process $|\overline{x}(t)|$ and the finiteness of the integral from time $\tau$ to $\tau'$. Let $\tau' \rightarrow +\infty$, the probability bound in \eqref{ineq: probability-bound} is 
\begin{align*}
{\rm Pr}\{\sup_{\tau \leq t} |\overline{x}(t)| \geq \epsilon'\} \leq \frac{c_1 |\overline{x}(0)|}{c_2 \epsilon'} e^{-c_{2}\tau} \leq \frac{c_1 |\overline{x}(0)|}{c_2 \epsilon'}
\end{align*}
Let $\epsilon:=\frac{c_1 |\overline{x}(0)|}{c_2 \epsilon'}$, then there indeed exists a function $\delta(\epsilon, \epsilon')=\frac{c_2 \epsilon'}{c_1}$ such that ${\rm Pr}\{\sup_{\tau \leq t} |\overline{x}(t)| \geq \epsilon'\} \leq \epsilon$ for any $|\overline{x}(0)| \leq \delta(\epsilon, \epsilon')$. Furthermore, since $\tau$ is arbitrarily chosen, then almost surely asymptotic stability defined in \eqref{ineq:as-as} holds. Taking $\tau \rightarrow \infty$ yields
\begin{align*}
\lim_{\tau \rightarrow \infty}{\rm Pr}\{\sup_{\tau \leq t} |\overline{x}(t)| \geq \epsilon'\} \leq \lim_{\tau \rightarrow \infty}\frac{c_1 |\overline{x}(0)|}{c_2 \epsilon'} e^{-c_{2}\tau}=0
\end{align*}
The proof is complete. 
\end{IEEEproof}

\begin{IEEEproof}[Proof of Theorem \ref{thm: ua-bounded}]
By Assumptions \ref{assumption: x} and \ref{assumption: e}, $\forall t > t_{k} \in \mathbb{R}_{+}, k \in \mathbb{Z}_{+}$, one has
\begin{align*}
\mathbb{E}(|\overline{x}(t)|) \leq \beta(\mathbb{E}(|\overline{x}(t_{k})|), t-t_{k}) + \overline{\chi}_{1}\mathbb{E}(|e|_{[t_{k}, t)}) + \chi_{2}(M)
\end{align*}	
with $|w|_{\mathcal{L}_{\infty}} \leq M$, and
\begin{align*}
\mathbb{E}(|e(t)|) \leq &\frac{w_{2}}{w_{1}}e^{L_{e}(t-t_{k})} G(|x(t_{k})|) \mathbb{E}(|e(t_{k})|) 
+\frac{L_{w}M}{w_{1}L_{e}}(e^{L_{e}(t-t_{k})}-1) \\
&+\frac{L_{x}}{w_{1}L_{e}}(e^{L_{e}(t-t_{k})}-1)\mathbb{E}(|x|_{[t_{k}, t)}) 
\end{align*}
Since under the self-triggered scheme in \eqref{TI}, the condition in \eqref{ineq: small-gain} assures that the  small-gain theorem holds for the interconnected system of $\mathbb{E}(|\overline{x}(t)|)$ and $\mathbb{E}(|e(t))$, the system with states $\overline{X}(t):=[\mathbb{E}(|\overline{x}(t)|); \mathbb{E}(|e(t))]$ is then input to state stable with respect to the external disturbance \cite{jiang2001input}. In particular, there exists a class $\mathcal{KL}$ function $\beta^{'}(\cdot, \cdot)$ and a class $\mathcal{K}$ function $\chi(\cdot)$ such that
\begin{align}
\label{ineq: iss}
|\overline{X}(t)|\leq \beta^{'}(|\overline{X}(0)|, t) + \chi(M).
\end{align}
From \eqref{ineq: iss}, $\forall |\overline{x}_{0}| \leq \Delta_0$, one knows that $\mathbb{E}\{|\overline{x}(t)|\} \leq \beta'(\Delta_{0}, 0)+\chi(M) \triangleq \epsilon(M, \Delta_0)$ and $\lim_{t \rightarrow +\infty}\mathbb{E}\{|\overline{x}(t)|\} \leq \chi(M)$. The proof is complete. 
\end{IEEEproof}
\begin{IEEEproof}[Proof of Theorem \ref{theorem:pc}]
Suppose the claim in Theorem \ref{thm: ua-bounded} holds with \eqref{ineq: iss}, for any given $\epsilon > 0$, the probability of the system state $\overline{x}$ exiting a given set $\Omega_{s}=\{\overline{x} \in \mathbb{R}^{n} | |\overline{x}|\leq \Delta\}$ at time $t$ can be bounded by
\begin{align}
{\rm Pr}\{|\overline{x}(t)| \geq \Delta + \epsilon \} &\leq \frac{\mathbb{E}(|\overline{x}(t)|)}{\Delta+\epsilon} \\
& \leq \frac{\beta^{'}(\mathbb{E}(|\overline{x}(0)|, t))+\chi(M)}{\Delta+\epsilon}
\label{ineq: probability-bound}
\end{align}
The first inequality follows by Markov's Inequality and the second inequality holds due to the \emph{input to state stability}. Taking the limit of time to infinity leads to
\begin{align}
\lim_{t \rightarrow +\infty}{\rm Pr}\{|\overline{x}| \geq \Delta+\epsilon \}\leq \frac{\chi(M)}{\Delta+\epsilon}.
\end{align}
Thus, the VNS in \eqref{shs} with bounded external disturbance is practically stable in probability with the probability bound $\epsilon(M, \Delta)=\frac{\chi(M)}{\Delta+\epsilon}$. The proof is complete. 
\end{IEEEproof}

\begin{IEEEproof}[Proof of Proposition \ref{prop: encoder-decoder}]
The proof is based on an induction method. Since we assume that the encoder and decoder share the initial value of $\hat{\overline{x}}_{0}^{+}$ and $U_{0}^{+}$ and the actual value of initial state $\overline{x}(0)$ lies in the hypercubic box with $\hat{\overline{x}}_{0}^{+}$ being its centroid and $2U_{0}^{+}$ being its edge length, the case of $k=0$ holds. Next, suppose the case of $k$ holds, i.e., the state $\overline{x}(t_{k})$ at time instant $t_k$ lies in the hypercubic box with parameters $(\hat{\overline{x}}_{k}^{+}, U_{k}^{+})$ and $|\overline{x}(t_k)-\hat{\overline{x}}_{k}^{+}| \leq U_{k}^{+}$. In the sequel, we show that the case of $(k+1)^{th}$ holds under the recursive equations \eqref{eq: U_k} and \eqref{eq: x_q}.

 First, consider the estimation error $e(t):=\overline{x}(t)-(\hat{x}-x^{d})=x(t)-\hat{x}$ over time interval $[t_{k}, t_{k+1}), \forall k \in \mathbb{Z}_{\geq 0}$. Let $t^{-}$ and $t^{+}$ denote the time instants before and after the bits are received respectively. By Assumption \ref{assumption: e}, one has
\begin{align}
|e(t)| \leq &\frac{w_{2}}{w_{1}}e^{L_{e}(t-t_{k})} |e(t_{k}^{+})| +\frac{L_{x}}{w_{1}L_{e}}(e^{L_{e}(t-t_{k})}-1)|\overline{x}|_{[t_{k}, t)} \nonumber \\
&+ \frac{L_{w}}{w_{1}L_{e}}(e^{L_{e}(t-t_{k})}-1)M
\label{ineq: |e|}
\end{align}
Similarly, Assumption \ref{assumption: x} leads to
\begin{align}
\label{ineq: |x|}
|\overline{x}|_{[t_{k}, t)} \leq \beta(|\overline{x}(t_{k})|, 0)+ \overline{\chi}_{1}|e|_{[t_{k}, t)} + \chi_{2}(M)
\end{align}
Substituting \eqref{ineq: |x|} into \eqref{ineq: |e|}  and letting $t=t_{k+1}^{-}$, since $|e(t_{k+1})|=|e(t)|_{[t_k, t)}$, one has
\begin{align*}
&\underbrace{\left(1-\frac{L_x\overline{\chi}_{1}}{w_{1}L_e}(e^{L_{e}T_{k}}-1)\right)}_{\eta_{k+1}}|e(t_{k+1}^{-})| \leq \frac{w_{2}}{w_{1}}e^{L_{e}T_{k}}|e(t_{k}^{+})| \\
&+ \frac{e^{L_{e}T_{k}}-1}{w_{1}L_{e}}\Big(L_{x}\beta(|\overline{x}(t_k)|, 0)+L_{w}M+L_{x}\chi_{2}(M)\Big)
\end{align*}
Since the transmission sequence $\{t_{k}\}$ is generated by \eqref{TI} and the small-gain condition \eqref{ineq: small-gain} holds, $\eta_{k} > 0, \forall k \in \mathbb{Z}_{+}$. Suppose $|e(t_{k}^{+})|\leq U_{k}^{+}$, then the following inequality
\begin{align}
\label{ineq: U_{k+1}}
|e(t_{k+1}^{-})|& \nonumber \\
 \leq & \frac{1}{\eta_{k+1}}\bigg(\frac{w_{2}}{w_{1}}e^{L_{e}T_{k}}U_{k}^{+}+\frac{e^{L_{e}T_{k}}-1}{w_{1}L_{e}}\Big(L_{x}\beta(|\hat{\overline{x}}_{k}^{+}|+U_{k}^{+}, 0) \nonumber \\&+L_{w}M+L_{x}\chi_{2}(M)\Big)\bigg):=U_{k+1}
\end{align}
holds due to $\eta_{k} > 0$ and $|\overline{x}(t_{k})| \leq |\hat{\overline{x}}_{k}^{+}|+U_{k}^{+}$. Upon successfully receiving $R_{k+1}$ blocks of bits at time $t_{k+1}^{+}$, one has $|e(t_{k+1}^{+})| \leq 2^{-R_{k+1}}|e(t_{k+1}^{-})|$. Let 
\begin{align*}
U_{k+1}^{+} :=& \frac{2^{-R_{k+1}}}{\eta_{k+1}}\bigg(\frac{w_{2}}{w_{1}}e^{L_{e}T_{k}}U_{k}^{+}+\frac{e^{L_{e}T_{k}}-1}{w_{1}L_{e}}\Big(L_{x}\beta(|\hat{\overline{x}}_{k}^{+}|+U_{k}^{+}, 0) \\
\quad &+L_{w}M+L_{x}\chi_{2}(M)\Big)\bigg)
\end{align*}
then $|e(t_{k+1}^{+})| \leq U_{k+1}^{+}$. Since the transmission time interval $[t_{k}, t_{k+1}]$ is selected arbitrarily, $\{U_{k}^{+}\}$ is a sequence of upper bounds on the estimation errors $\{e(t_{k}^{+})\}$, i.e., $|\overline{x}(t_k)-\hat{\overline{x}}_{k}^{+}| \leq U_{k}^{+}, \forall k \in \mathbb{Z}_{\geq 0}$.

Secondly,  the state estimate $\hat{\overline{x}}_{k+1}^{+}$ is updated by selecting the centroid of an updated hypercubic box that contains $\overline{x}(t_{k+1})$. To be specific, during the time interval $[t_k, t_{k+1})$, the centroid $\hat{\overline{x}}(t)$ of the hypercubic box is updated by both encoder and decoder according to the dynamic equation $\dot{\hat{\overline{x}}}=f_{\overline{x}}(\overline{x}, 0, 0)$ with initial value $\hat{\overline{x}}_{k}^{+}$. The centroid of the expanded hypercubic box at time instant $t_{k+1}^{-}$ before receiving new information bits, is $\Phi(\hat{\overline{x}}(t_k), T_k):=\hat{\overline{x}}(t_{k+1}^{-})=\hat{\overline{x}}_{k}^{+}+\int_{t_{k}}^{t_{k}+T_k} f_{\overline{x}}(\hat{\overline{x}}, 0, 0)dt$. By inequality \eqref{ineq: U_{k+1}}, one knows that the state $\overline{x}(t_{k+1})$ is guaranteed to lie in an expanded hypercubic box with the centroid $\hat{\overline{x}}(t_{k+1}^{-})$ and the size $U_{k+1}$. Upon receiving $R_{k+1}$ blocks of bits at time instant $t_{k+1}$, the expanded hypercubic box is partitioned into $2^{nR_{k+1}}$ number of sub-boxes with each sub-box's centroid being encoded by a binary sequence $\{b_{j}\}_{j=1}^{R_{k+1}}$.  Thus, for a given centroid $\Phi(\hat{\overline{x}}(t_k), T_k)$, a given box length $U_{k+1}$ and $\{b_{j}\}_{j=1}^{R_{k+1}}$, the function $q(b_{ji}) \in \{-1, 1\}$ decodes the $i^{th}$ bit in the $j^{th}$ block as a relative "position" to the centroid $\Phi(\hat{\overline{x}}(t_k), T_k)$. By a uniform quantization method \cite{martins2006feedback}, the centroid of the sub-box that contains the actual state $\overline{x}(t_{k+1})$ is thus 
\begin{align*}
\hat{\overline{x}}_{k+1}^{+}=U_{k+1}\sum_{j=1}^{R_{k+1}}\frac{1}{2^j}q\big(b_{j}(k+1)\big)
+\Phi(\hat{\overline{x}}_{k}^{+},T_{k})
\end{align*}
 
The proof is complete. 
\end{IEEEproof}

	\begin{IEEEbiography}[]{Bin Hu}
	received the B.S. degree in
	automation from Hefei University of Technology,
	Hefei, China, in 2007, the M.S. degree in control
	and system engineering from Zhejiang University,
	Hangzhou, China, in 2010, and the Ph.D. degree
	in electrical engineering from the University of
	Notre Dame, Notre Dame, IN, USA in 2016. He is currently with the department of engineering technology at Old Dominion University in Norfolk, VA, USA. His research interests include stochastic networked control systems, information theory, switched control systems,
	distributed control and optimization, and human machine interaction.
\end{IEEEbiography}

	\begin{IEEEbiography}[]{Michael D. Lemmon}
 (SM’15) received the B.S.
	degree in electrical engineering from Stanford University,
	Stanford, CA, USA, in 1979 and the Ph.D.
	degree in electrical and computer engineering from
	Carnegie-Mellon University, Pittsburgh, PA, USA,
	in 1990.
	
	He is a Professor of electrical engineering at the
	University of Notre Dame, Notre Dame, IN, USA.
	His work has been funded by a variety of state and
	federal agencies that include the National Science
	Foundation, Army Research Office, Defense Advanced
	Research Project Agency, and Indiana’s 21st Century Technology Fund.
	His research deals with real-time networked control systems with an emphasis
	on understanding the impact that reduced feedback information has on overall
	system performance.
	
	Dr. Lemmon was an Associate Editor for the IEEE TRANSACTIONS ON
	NEURAL NETWORKS and the IEEE TRANSACTIONS ON CONTROL SYSTEMS
	TECHNOLOGY. He chaired the first IEEE working group on hybrid
	dynamical systems and was the program chair for a hybrid systems workshop
	in 1997. Most recently, he helped forge a consortium of academic, private
	and public sector partners to build one of the first metropolitan scale sensoractuator
	networks (CSOnet) used in monitoring and controlling combinedsewer
	overflow.
\end{IEEEbiography}
\end{document}